\documentclass[11pt,a4paper]{article}
 
\pdfoutput=1 

\usepackage{jheppub} 

\usepackage[T1]{fontenc} 
\usepackage{amsmath}
\usepackage{enumitem}
\usepackage{xcolor, xspace, graphicx}
\usepackage{amssymb,amsthm,cancel,hyperref}
\usepackage{mathrsfs,wasysym}
\usepackage{booktabs}
\usepackage{subcaption}
\usepackage{tikz}
\usetikzlibrary{calc}
\usepackage{placeins}
\usepackage{tabu}
\usepackage{pdflscape}
\usepackage{afterpage}
\usepackage{makecell}

\newcommand{\lsim}{\lesssim}

\let\originalleft\left
\let\originalright\right
\renewcommand{\left}{\mathopen{}\mathclose\bgroup\originalleft}
\renewcommand{\right}{\aftergroup\egroup\originalright}

\newcommand{\zc}{z_{\rm cut}}

\newcommand{\la}{\lambda_\alpha}

\newcommand{\sherpa}{S\protect\scalebox{0.8}{HERPA}\xspace}
\newcommand{\pythia}{P\protect\scalebox{0.8}{YTHIA}\xspace}
\newcommand{\jewel}{J\protect\scalebox{0.8}{EWEL}\xspace}
\newcommand{\herwig}{H\protect\scalebox{0.8}{ERWIG}\xspace}

\newcommand{\comix}{C\protect\scalebox{0.8}{OMIX}\xspace}

\newcommand{\Caesar}{C\protect\scalebox{0.8}{AESAR}\xspace}
\newcommand{\fastjet}{F\protect\scalebox{0.8}{AST}J\protect\scalebox{0.8}{ET}\xspace}
\newcommand{\rivet}{R\protect\scalebox{0.8}{IVET}\xspace}
\newcommand{\recola}{R\protect\scalebox{0.8}{ECOLA}\xspace}
\newcommand{\collier}{C\protect\scalebox{0.8}{OLLIER}\xspace}
\newcommand{\OpenLoops}{O\protect\scalebox{0.8}{PEN}L\protect\scalebox{0.8}{OOPS}\xspace}

\newcommand{\softdrop}{\textsf{SoftDrop }}
\newcommand{\fjcontrib}{\textsf{fjcontrib}}

\newcommand{\muR}{\ensuremath{\mu_{\text{R}}}}
\newcommand{\muF}{\ensuremath{\mu_{\text{F}}}}

\newcommand{\zcut}{\ensuremath{z_{\text{cut}}}}

\newcommand{\LO}{\text{LO}\xspace}
\newcommand{\NLO}{\text{NLO}\xspace}    
\newcommand{\NLL}{\text{NLL}\xspace}

\newcommand{\NLOpNLL}{\ensuremath{\NLO+\NLL}\xspace}
\newcommand{\LOpNLL}{\ensuremath{\LO+\NLL}\xspace}    
\newcommand{\NLOpNLLp}{\ensuremath{\NLOpNLL^\prime}\xspace}
\newcommand{\LOpNLLp}{\ensuremath{\LOpNLL^\prime}\xspace}

\def\beq{\begin{equation}}  
\def\eeq{\end{equation}}
\def\({\left(}
\def\){\right)}
\def\[{\left[}
\def\]{\right]}

\usepackage{color}
\definecolor{darkblue}{rgb}{0,0,0.5}
\definecolor{darkred}{rgb}{0.5,0,0}
\definecolor{darkgreen}{rgb}{0,0.5,0}

\title{Jet angularities in dijet production in proton--proton
and heavy-ion collisions at RHIC}

\author[a, b, c]{Yang-Ting Chien,} 

\author[a, b, c]{Oleh Fedkevych,} 

\author[d]{Daniel Reichelt,} 

\author[e]{Steffen Schumann} 

\affiliation[a]{Physics and Astronomy Department, Georgia State University, Atlanta, GA 30303, USA}

\affiliation[b]{Center for Frontiers in Nuclear Science, Stony Brook University, Stony Brook, NY 11794, USA}

\affiliation[c]{Theory Center, Jefferson Lab, Newport News, Virginia 23606, USA}

\affiliation[d]{Institute for Particle Physics Phenomenology, Department of Physics, Durham University, Durham DH1 3LE, United Kingdom}

\affiliation[e]{Institut f\"ur Theoretische Physik, Georg-August-Universit\"at G\"ottingen, Friedrich-Hund-Platz 1, 37077, G\"ottingen, Germany}

\emailAdd{ytchien@gsu.edu}
\emailAdd{ofedkevych@gsu.edu}
\emailAdd{daniel.reichelt@durham.ac.uk}
\emailAdd{steffen.schumann@phys.uni-goettingen.de}

\preprint{MCNET-24-04 IPPP/24/14, JLAB-THY-24-4010}
\abstract
{
  We study jet  angularities for dijet production at the Relativistic Heavy Ion Collider (RHIC) in proton--proton ($\rm pp$)
  and nucleus--nucleus (AA) collisions at $200$ GeV nucleon--nucleon center-of-mass collision energy.
In particular, we provide $\mathrm{NLL}$ resummed predictions for angularity observables of groomed and ungroomed jets produced in $\rm pp$ collisions matched to next-to-leading order QCD calculations
resulting in $\mathrm{NLO+NLL^\prime}$ accuracy. 
Our parton-level predictions are corrected for non-perturbative effects, such as hadronization and underlying event, using parton-to-hadron level transfer matrices obtained with the \sherpa event generator.
Furthermore, we use the Q-\pythia and \jewel generators to estimate the impact of the interaction between quarks and gluons
produced by the parton shower with the dense medium formed in heavy-ion collisions on the considered jet angularities.
}
\begin{document}
\maketitle
\section{Introduction}
Jets are collimated sprays of hadronic particles abundantly produced in collider experiments. 
In a variety of different production processes jets can be instrumented for tests of fundamental properties
of the Standard Model (SM) of particle physics as well as for searches of new particles predicted by various SM extensions.
As a consequence, within the last decade jet substructure physics has received a lot of attention from both theoretical and experimental communities. 
Due to the efforts of numerous research groups the 
field quickly matured and many crucial applications were identified. 
For example, jet substructure was used in fits of the strong coupling constant~\cite{Britzger:2017maj, CMS:2013vbb, ATLAS:2017qir, ATLAS:2015yaa, CMS:2014mna}, to constrain parton distribution functions (PDFs)~\cite{ATLAS:2021qnl, ATLAS:2013pbc, CMS:2014qtp, CMS:2016lna, AbdulKhalek:2020jut, Harland-Lang:2017ytb, Pumplin:2009nk, Watt:2013oha}, to test high-precision perturbative QCD calculations~\cite{CMS:2021iwu, ALICE:2021njq}, or to search for heavy hadronically decaying resonances predicted by various beyond SM theories~\cite{Soper:2010xk, Godbole:2014cfa, Chen:2014dma, Adams:2015hiv}.
Furthermore, jet substructure observables are actively used as input variables for machine learning algorithms, see \textit{e.g.} Refs.~\cite{Larkoski:2017jix,Kasieczka:2019dbj,Benato:2020sbi}, and for particle tagging purposes~\cite{Fedkevych:2022mid, Caletti:2021ysv, Dreyer:2021hhr, Cavallini:2021vot, Khosa:2021cyk, Dreyer:2020brq,Baron:2023hkp}.
Given that the substructure of jets is sensitive to interactions between partons produced in a hard scattering process with the dense medium formed in proton--nucleus ($\rm pA$) or nucleus--nucleus ($\rm AA$) collisions, see \textit{e.g.} Refs.~\cite{Lapidus:2017dek,Zapp:2017ria,Tywoniuk:2017dzi,Casalderrey-Solana:2017mjg, Casalderrey-Solana:2019ubu, Mangano:2017plv,Qin:2017roz,Milhano:2017nzm,Chang:2017gkt,KunnawalkamElayavalli:2017hxo}, it provides an important tool to study quark-gluon plasma (QGP) signatures.

Whereas many  measurements of jet substructure observables were performed at the Large Hadron Collider (LHC)~\cite{ALICE:2017nij, ALICE:2018dxf,  CMS:2011hzb, CMS:2013lhm,  CMS:2013kfv,  CMS:2013lua, CMS:2014qvs,  CMS:2016php, CMS:2017eqd, CMS:2017pcy,  CMS:2018mqn, CMS:2018fof,  CMS:2018vzn, CMS:2018ypj, CMS:2019fak,  ATLAS:2011kzm, ATLAS:2012nnf, ATLAS:2012am, ATLAS:2014lzu,  ATLAS:2017nre, ATLAS:2017zda, ATLAS:2017xqp, ATLAS:2018jsv, ATLAS:2019dty, ATLAS:2019mgf}, the number of such analyses at the Relativistic Heavy Ion Collider (RHIC) is much smaller, see \textit{e.g.} Refs.~\cite{Kauder:2017mhg, STAR:2020ejj, STAR:2021lvw, Connors:2017ptx}. 
However, due to the significantly different collision energies, considerable differences in the jet substructure can be anticipated for jets produced at
RHIC with respect to those measured at LHC.
In particular, with the decrease of collision energy and, as a consequence, of the momentum transfer one would expect a much larger impact of non-perturbative physics~\cite{Moraes:2007rq}.
This, in turn, challenges the non-perturbative models as implemented in general-purpose Monte Carlo event generators such as \pythia~\citep{Bierlich:2022pfr,Sjostrand:2006za}, \herwig~\citep{Bellm:2019zci,Corcella:2000bw} and \sherpa~\citep{Sherpa:2019gpd,Gleisberg:2008ta} and may require a dedicated re-tuning of model parameters, see \cite{Aguilar:2021sfa, Reichelt:2021svh}.
Therefore, jet substructure studies at RHIC require not only precise first-principles theoretical predictions but also a careful consideration of non-perturbative effects which, as it was demonstrated in Ref.~\cite{Reichelt:2021svh}, may lead to significant bin-migration processes seriously affecting the shape of jet substructure observables. 
Given $\rm pp$ collisions are used as a baseline to compare against $\rm pA$ and $\rm AA$ results in QGP searches, a solid understanding of jet substructure physics at RHIC not only in $\rm pA$ and $\rm AA$ but also in $\rm pp$ collisions is needed to facilitate reliable interpretation of experimental measurements. 

In this paper we consider a set of jet substructure observables called jet angularities~\cite{Berger:2003iw,Almeida:2008yp,Larkoski:2014pca} which were measured for jets produced in $\rm pp$ collisions at the LHC by the ALICE~\cite{ALICE:2021njq}, ATLAS~\cite{ATLAS:2012nnf} and CMS~\cite{CMS:2018ypj, CMS:2021iwu} experiments.
Additionally, the ALICE collaboration measured  angularities for jets emerging from $\rm AA$ collisions~\cite{ALICE:2018dxf}.
A closely related  set of observables (jet mass and jet shape) was also measured by CDF~\cite{CDF:2011loy} for $p\bar{p}$ collisions and  later by ATLAS, CMS
and ALICE in $\rm pp$~\cite{CMS:2013lhm,  CMS:2013kfv,  CMS:2017pcy,  CMS:2018fof,  CMS:2018mqn, CMS:2018vzn, CMS:2018ypj, CMS:2019fak,  ATLAS:2012nnf, ATLAS:2012am, ATLAS:2017nre, ATLAS:2017zda, ATLAS:2018jsv,  ATLAS:2019dty, ATLAS:2019mgf} and $\rm AA$ collisions~\cite{ALICE:2017nij, ALICE:2018dxf,  CMS:2013lhm, CMS:2018mqn, CMS:2018fof,  ATLAS:2017nre, ATLAS:2018jsv}, respectively.
Additionally, the ALICE collaboration measured the jet mass for jets produced in $\rm pA$ scattering~\cite{ALICE:2017nij}.

Important progress has been achieved in recent years on the theoretical description of jet substructure observables. For an overview see~\cite{Marzani:2019hun}. 
Specifically, high-precision results for jet angularities were presented in Refs.~\cite{Caletti:2021oor, Reichelt:2021svh, Kang:2018qra, Kang:2018vgn, Almeida:2014uva, Dasgupta:2022fim, Budhraja:2023rgo}.
In particular, the  $\mathrm{NLO+NLL^\prime}$ accurate jet angularity
predictions from Refs.~\cite{Caletti:2021oor, Reichelt:2021svh}
were compiled using the resummation plugin to the \sherpa generator
framework~\cite{Gerwick:2014gya,Baberuxki:2019ifp} and hence are largely
automated.
The main scope of this paper is to provide $\mathrm{NLO+NLL^\prime}$ predictions for future measurements of jet angularities in $\rm pp$
collisions at RHIC, which is currently operating the new sPHENIX detector in data-taking mode~\cite{Okawa:2023asr}, as well as to estimate
the r\^{o}le of non-perturbative physics and medium-induced effects through dedicated Monte Carlo simulations.

This paper is organized as follows: in Section~\ref{sec:def} we provide definitions of the observables under consideration and briefly discuss the details of our theoretical computations, in Section~\ref{sec:resum} we present resummed predictions for jet angularities at \NLOpNLLp accuracy level, in Section~\ref{sec:np_effects} we correct our resummed predictions for non-perturbative effects using parton-to-hadron level transfer matrices as introduced in Ref.~\cite{Reichelt:2021svh} and compare the results against predictions from the \sherpa and \pythia event generators.
Finally, in Section~\ref{sec:medium_effects} we estimate the impact of medium effects based on simulations with the Q-\pythia code~\cite{Armesto:2009fj}, derived from the \pythia6 generator~\cite{Sjostrand:2006za}, as well as \jewel~\cite{Zapp:2013vla}.  
Our conclusions and a discussion of possible further steps are presented in Section~\ref{sec:conclusions}.

\section{Event selections and observable definitions}\label{sec:def}
We consider dijet final states produced in $\rm pp$  collisions at
$\sqrt{s_{\rm pp}} = 200$ GeV  center-of-mass collision energy. 
We use the \fastjet~\cite{Cacciari:2011ma} code to cluster final-state particles into jets according to the anti-$k_t$ jet algorithm~\cite{Cacciari:2008gp} with radius parameter $R_0 = 0.4$ and standard $E$-scheme recombination,
\emph{i.e.}\ the momenta of particles that get combined are simply added such that the resulting jet momentum is given by the sum of its constituents' momenta. 
We require at least two jets that satisfy
\begin{equation}
  p_{T,j_1}> 30\,\text{GeV}\,,\quad   p_{T,j_2}> 20\,\text{GeV} \,
  \quad\text{and}\quad
  |y_{j_1,j_2}|<0.7\,,
\end{equation}
which is in accordance with the RHIC detector acceptance characteristics and the sPHENIX beam
use proposal~\cite{Okawa:2023asr}.
Through the staggered transverse momentum cuts we avoid perturbative instabilities
appearing for exactly balanced back-to-back dijet kinematics, see Ref.~\cite{Currie:2018xkj}
for a detailed discussion.
In each selected event, we consider the two leading transverse momentum jets,
and for each jet individually evaluate the angularity observables
~\cite{Berger:2003iw,Almeida:2008yp,Larkoski:2014pca} defined as
\begin{equation}\label{eq:ang-def}
\la^\kappa =
\sum_{i \in \text{jet}}\left(\frac{p_{T,i}}{\sum_{j \in \text{jet}} p_{T,j}}\right)^\kappa\left(\frac{\Delta_i}{R_0} \right)^\alpha\, \equiv
\sum_{i \in \text{jet}}z^\kappa_i\left(\frac{\Delta_i}{R_0} \right)^\alpha\,,
\end{equation}
where
\begin{equation}\label{eq:dist-def}
\Delta_i=\sqrt{(y_i-y_\text{jet})^2+(\phi_i-\phi_\text{jet})^2}\,,
\end{equation}
is the standard Euclidean azimuth-rapidity distance between particle $i$ and the jet axis.
The concept of infrared and collinear (IRC) safety requires $\kappa=1$ and $\alpha>0$.
Therefore, we limit ourselves to three commonly studied cases: $\lambda^1_{1/2}$ (Les Houches angularity or LHA), $\lambda^1_{1}$ (Jet Width) and $\lambda^1_{2}$ (Jet Thrust)~\cite{Larkoski:2014pca, CMS:2021iwu, Andersen:2016qtm}.
Also, in order to keep our notation simple, in the rest of the paper we omit the subscript $\kappa$ assuming $\kappa = 1$.
Ultimately, we consider the sum of the distributions individually determined for
the two leading jets as described above.

Angularities with $\alpha \le 1$ are sensitive to recoil against soft emissions~\cite{Banfi:2004yd,Larkoski:2013eya},
leading to a rather complicated resummation structure.
To avoid the complication caused by recoil effects, for $\lambda_{1/2}$ and $\lambda_{1}$  we evaluate the
distance measure in Eq.~\eqref{eq:dist-def} with respect to the jet axis obtained by reclustering the jet constituents
with the anti-$k_t$ algorithm but using the Winner-Take-All (WTA) recombination scheme~\cite{Larkoski:2014uqa}.

We also consider groomed jets where as a groomer we use the \softdrop  algorithm with parameters $\zc = 0.2$ and
$\beta = 0$~\cite{Marzani:2017mva, Larkoski:2014wba} and the Cambridge--Aachen (C/A) algorithm \cite{Dokshitzer:1997in, Wobisch:1998wt}
for jet reclustering. 
To this end we use the \fjcontrib~implementation of \softdrop in the \fastjet package.
The angularity is then computed on the resulting groomed jet, \textit{i.e.}\ both sums in Eq.~\eqref{eq:ang-def} are restricted to the particles
that survived the grooming.
Also for groomed jets we adopt the WTA prescription for angularities with $\alpha \le 1$.

In practice, we use the \rivet framework~\cite{Buckley:2010ar,Bierlich:2019rhm}
to perform the Monte Carlo analysis and use Matplotlib~\cite{Hunter:2007ouj} for
plotting.

\section{All-order resummation at next-to-leading logarithmic accuracy}\label{sec:resum}

To perform NLL resummation for jet angularities we use the
\sherpa implementation of the \Caesar resummation formalism~\citep{Banfi:2004yd,Banfi:2010xy}, first presented in~\citep{Gerwick:2014gya}. 
This framework has been  employed to obtain resummed predictions for \softdrop thrust~\citep{Marzani:2019evv} and multijet resolution scales~\citep{Baberuxki:2019ifp} in electron--positron collisions, as well as \NLOpNLLp predictions for \softdrop groomed hadronic event shapes~\citep{Baron:2020xoi}, for 
jet angularities in dijet and $Z$+jet production at the
LHC~\cite{Caletti:2021oor, Caletti:2021ysv, Reichelt:2021svh} and, recently,
for plain and groomed event shapes in neutral-current deep inelastic scattering~\cite{Knobbe:2023ehi,H1:2024aze,H1:2024pvu}, as well as event-shape observables in hadronic Higgs-boson decays~\cite{Gehrmann-DeRidder:2024avt}.

For an applicable jet-substructure observable, the all-order cumulative cross
section for observable values up to $v$, with $L=-\ln(v)$, can be written as a
sum over partonic channels $\delta$:
\begin{equation}\label{eq:CAESAR}
  \begin{split}
    \Sigma_\mathrm{res}(v) &= \sum_\delta \Sigma_\mathrm{res}^\delta(v)\,,\,\,\text{with} \\  
    \Sigma_\mathrm{res}^\delta(v) &= \int d\mathcal{B_\delta}
    \frac{\mathop{d\sigma_\delta}}{\mathop{d\mathcal{B_\delta}}} \exp\left[-\sum_{l\in\delta}
      R_l^\mathcal{B_\delta}(L)\right]\mathcal{S}^\mathcal{B_\delta}(L)\mathcal{F}^\mathcal{B_\delta}(L)\mathcal{H}^{\delta}(\mathcal{B_\delta})\,,
  \end{split}
\end{equation}
where $\frac{\mathop{d\sigma_\delta}}{\mathop{d\mathcal{B_\delta}}}$ is the
fully differential Born cross section for the subprocess $\delta$ and
$\mathcal{H}$ implements constraints on the Born phase space $\mathcal{B}$.
The label $\mathcal{F}$ represents the multiple emission function which,
for additive observables such as the angularities considered in this paper,
is simply given by $\mathcal{F}(L) = e^{-\gamma_E R^\prime}/\Gamma(1+R^\prime)$,
with $R^\prime(L)=\partial R/\partial L$ and
$R(L)=\sum_{l\in \delta} R_l(L)$.
The soft function $\mathcal{S}$ implements the non-trivial aspects of color
evolution. In our notation this includes the effect of non-global
logarithms \cite{Dasgupta:2001sh}. Since the relevant functions depend on the jet radius only, we here
can use the expressions extracted numerically in the leading-color approximation
from \cite{Reichelt:2021svh}, based on the code from \cite{Caletti:2021oor}.
The collinear radiators $R_l$ for the hard legs $l$ are summarized
in~\citep{Banfi:2004yd} for a general global observable $V$ scaling for the
emission of a soft-gluon of relative transverse momentum $k_t^{(l)}$,
rapidity $\eta^{(l)}$ and azimuthal angle $\phi^{(l)}$ with respect to leg $l$ as
\begin{equation}\label{eq:CAESAR_param}
  V(k)=\left(\frac{k_{t}^{\left(l\right)}}{\mu_Q}\right)^{a}e^{-b_{l}\eta^{\left(l\right)}}d_{l}\left(\mu_Q\right)g_{l}\left(\phi^{(l)}\right)\,,
\end{equation}
while here we have to take into account the modifications from
\cite{Dasgupta:2012hg, Caletti:2021oor} to account for the phase-space
restrictions implied by the finite jet radius. For the jet angularities defined
by Eq.~\eqref{eq:ang-def} it holds \cite{Caletti:2021oor}
\begin{equation}\label{eq:ang_parameter}
  a=1\,,\quad b_l=b=\alpha-1\,,\quad
  g_l d_l=\left(\frac{2\cosh \eta_\text{jet}}{R_0}\right)^{\alpha-1} \frac{\mu_Q}{p_{T,\text{jet}} R_0}\,.
\end{equation}
Here, we denote the transverse momentum and rapidity (noting that the
resummation is performed around configurations corresponding to massless jets)
by $\eta_\text{jet},\; p_{T,\text{jet}}$. We choose $\mu_Q = p_{T,\text{jet}} R_0$,
simplifying the last term in $g_l d_l$ to unity.

Generalized expressions for the radiator functions, that we use here when considering
\softdrop groomed final states with general parameters $\zcut$ and
$\beta$, have been presented in~\cite{Baron:2020xoi}. As discussed in
\cite{Caletti:2021oor, Reichelt:2021svh} our resummation is strictly valid in
the limit of small $\zcut$, \textit{i.e.} $\lambda_\alpha \ll \zcut \ll
1$. However, for $\beta=0$ finite-$\zcut$ corrections are already present at the
(leading) single-logarithmic accuracy~\cite{Dasgupta:2013ihk}, and
have been found to have a negligible effect in practice~\cite{Marzani:2017mva}
at least for $\zcut=0.1$. We will assume that similar statements can be made for
$\zcut=0.2$ and do not include any related corrections.

The \Caesar resummation plugin to \sherpa hooks into the event generation
framework, facilitating the process management, and providing access to the
\comix matrix-element generator~\citep{Gleisberg:2008fv}, as well as phase-space
integration and event-analysis functionalities. The \sherpa framework is also
used to compile all the required higher-order tree-level and one-loop
calculations. The plugin implements the building blocks of the \Caesar master formula
Eq.~\eqref{eq:CAESAR}, along with the necessary expansion in the strong coupling
constant $\alpha_s$ used in the matching with fixed-order calculations. All
elements are evaluated fully differentially for each Born-level configuration
$\mathcal{B}_\delta$ of a given flavor and momentum configuration.

The treatment of the kinematic endpoint is implemented in the same way as in
Ref.~\citep{Baron:2020xoi} by shifting the relevant logarithms and adding
power-suppressed terms to achieve a cumulative distribution that approaches one
at the kinematic endpoint and has a smooth derivative approaching zero. In
particular, we introduce the additional parameters $p, x_L, v_\mathrm{max}$ and
modify all logarithms according to
\begin{equation}\label{eq:log_shift}
   \ln \left(\frac{x_L}{v}\right) \to \frac{1}{p} \ln\left[\left(\frac{x_L}{
         v}\right)^p-\left(\frac{x_L}{v_\mathrm{max}}\right)^p+1\right]=L\,.
\end{equation}
Here we set $v_\mathrm{max}$ to the numerically determined endpoint of the observable distribution
evaluated at NLO, and by default use $p=1$. We set $x_L=1$, while a variation of
this parameter can be used to assess uncertainties of the resummed
calculation. Note that $x_L\neq 1$ introduces additional NLL terms that need to
be subtracted.

To adequately describe intermediate and large angularity observable values,
the resummed calculation needs to be matched to an appropriate fixed-order
calculation. We use the \comix matrix-element generator shipped with
\sherpa to perform an NLO calculation for the dijet final state. In order to
capture the next-order corrections for jet shapes like the angularities, it is
sufficient to perform an NLO calculation for the 3-parton final state, with a
small cut-off $\lambda_\alpha^\text{min}$ in the observable to regularize the divergences
from double soft and triple collinear splittings. In practice we choose
$\lambda_\alpha^\text{min}$ to be smaller than the smallest bin edge considered,
and enforce the physical condition $\Sigma(0)=0$ in our matching
prescription. Note that this means that our calculation technically becomes
observable specific and we perform separate calculations for angularities
measured on the more central and more forward of the two leading
jets, while keeping the phase space for radiation in the other jet unrestricted.
We use the \sherpa implementation of the Catani--Seymour dipole
subtraction scheme~\citep{Gleisberg:2007md} and the interfaces to the
\recola~\citep{Actis:2016mpe,Biedermann:2017yoi} and
\OpenLoops~\citep{Cascioli:2011va} one-loop amplitude generators. 

To combine our NLL results with the fixed-order calculation, we perform a
multiplicative matching, separately for quark and gluon jets in order to ensure
an effective extraction of the $C_1$ coefficients necessary to achieve
\NLOpNLLp accuracy. While several more sophisticated jet-flavor algorithms have
appeared since then
\cite{Caletti:2022glq,Caletti:2022hnc,Czakon:2022wam,Gauld:2022lem,Caola:2023wpj},
we stick with the iterative application of \cite{Banfi:2006hf}
introduced in \cite{Caletti:2021oor} since it is sufficient for our
purpose. We stress again that we can perform all required fixed-order
calculations efficiently with the generic tools available in \sherpa to the
numerical accuracy necessary for an effective matching. After matching, we stack
the two distributions for the central and forward jet, to obtain a prediction
for the summed distribution that would be obtained by filling a histogram with
the values of both the leading jets for each event.

As parton density distributions we use the NNLO PDF4LHC21\_40\_pdfas set~\cite{PDF4LHCWorkingGroup:2022cjn}
with $\alpha_s(M_Z) = 0.118$, accessed via the LHAPDF library~\cite{Buckley:2014ana}. The central values for the perturbative factorization ($\mu_{\text{F}}$) and
renormalization ($\mu_{\text{R}}$) scales entering the calculation are set to
\begin{eqnarray}
  \mu_{\text{F}}=\mu_{\text{R}}=H_T/2\,.\label{eq:scalesjj}
\end{eqnarray}
To estimate the perturbative uncertainties of our predictions, we perform
on-the-fly~\cite{Bothmann:2016nao} $7$-point variations using
\begin{equation}\label{eq:7pointvar}
  \left\{
  	(\tfrac{1}{2}\muR, \tfrac{1}{2}\muF), 
  	(\tfrac{1}{2}\muR,\muF), 
  	(\muR,\tfrac{1}{2}\muF), 
  	(\muR,\muF), 
  	(\muR,2\muF),
  	(2\muR,\muF), 
  	(2\muR,2\muF)
  \right\}\,,
\end{equation}
as well as an independent variation of the parameter $x_L$ introduced in Eq.~\eqref{eq:log_shift}, corresponding to $x_L=1/2$ and $x_L=2$.

\begin{figure}
  \centering
  \includegraphics[width=1.0\linewidth]{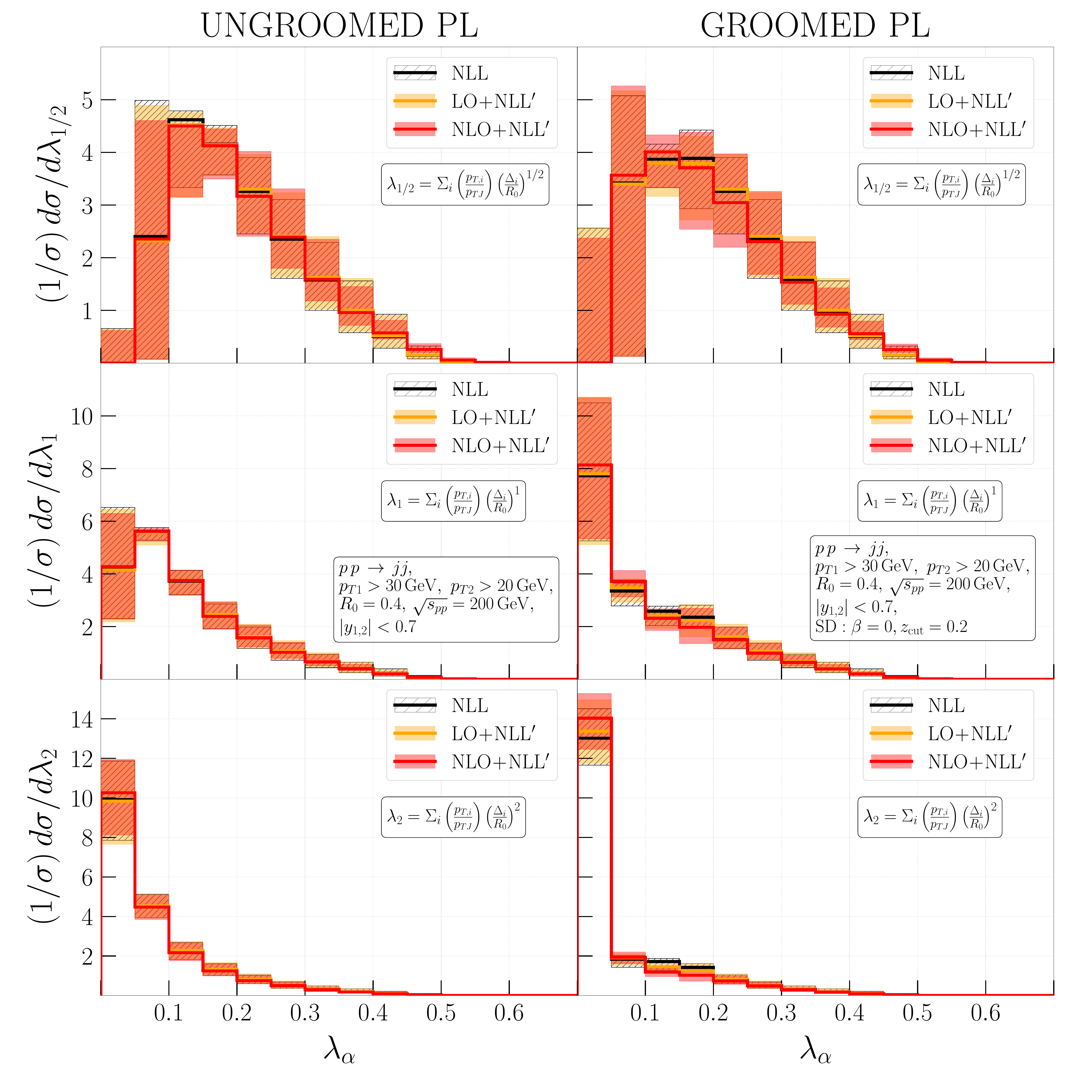}
  \centering
  \caption{Normalized parton-level resummed predictions for the $\lambda_{1/2}$, $\lambda_{1}$,
    $\lambda_{2}$ angularity distributions (top to bottom) without and with \softdrop grooming
    (left and right). We illustrate the effect of matching to the
    fixed-order calculation by showing the pure NLL result (red) compared to
    the leading-order matched one (orange) and our final \NLOpNLLp accurate matched
    result (black).}
  \label{fig:ang_all_pl}
\end{figure}

In Fig.~\ref{fig:ang_all_pl} we present resummed
predictions, without any non-perturbative correction dubbed parton level (PL),
for the jet-angularity observables in proton--proton collisions at
$\sqrt{s_{\rm pp}}=200\;\text{GeV}$.
In general, larger values of the angularities correspond to jets with increased
internal activity, either in the form of more or harder constituents. On the
other hand, the $\lambda_\alpha\to0$ limit corresponds to jets with no or only
soft and/or collinear radiation.  However, due to the different scaling of the
observables, equal values of the angularities across our choices for $\alpha$
probe different overall levels of softness. For this reason, $\lambda_{1/2}$
showcases the typical behavior of a jet shape, with a Sudakov peak produced by
logarithmic terms at all orders that dampen the rise present in the
leading-order contributions when $\lambda\to 0$.
The region to the left of the peak is fully resolved and we can follow the
suppression of events at very small $\lambda_{1/2}$ until the height of the first
bin is effectively $0$. This picture is much less clear for $\lambda_1$, since
the same values of $\lambda_1$ correspond to harder configurations, so the peak
is at smaller values and we can only observe the maximum of the distribution in
the bin corresponding to the second smallest $\lambda_1$ range. Finally, for
$\lambda_2$ this becomes even stronger and we do not observe any peak
structure given this binning choice. Of course the resummed distribution still approaches $0$ at
$\lambda=0$, however, the maximum would only be in the first bin, such that we
can not resolve the Sudakov peak structure. When comparing the ungroomed
distributions with the corresponding groomed ones, the latter, as expected, are
shifted towards smaller angularity values. \softdrop grooming eliminates part of
the jet constituents, thus sharpens the jet profile. While these differences are
of course to some extend an artifact of our chosen binning, this binning is
motivated by the resolution achieved by the LHC experiments, in particular the
measurement in \cite{CMS:2021iwu}, where a similar effect when changing $\alpha$
is observed. While from a theoretical point of view it might be interesting to
study the deep IR limit, see for example \cite{Caletti:2021oor}, we here stick
to predictions in a setting that should roughly be achievable in an actual 
experiment at RHIC.

We only find moderate effects from the matching of resummation and fixed-order
calculations at LO and NLO. When going from the pure NLL results to
the leading order accurate \LOpNLLp, all distributions are slightly shifted towards
larger values of the respective angularities. This is most visible in the groomed case, where for
$\alpha=1,2$ the height of the first bin is visibly reduced. For $\alpha=1/2$ the
first bin is already suppressed, but we still observe a small shift in the
peak position. The changes in the ungroomed distributions are less prominent but
qualitatively consistent with the above observations. The effect in some cases
becomes stronger after including NLO corrections, which are small in all
cases.

However, the impact of including fixed-order corrections, even at LO, on
the central value is dwarfed by the remaining uncertainty. These are
estimated by the envelope of \mbox{$7$-point} variations of the renormalization and factorization
scales $\mu_R$, $\mu_F$, and variations of the resummation scale factor $x_L$ introduced
in Eq.~\eqref{eq:log_shift}. The widths of the uncertainty bands are dominated by the
$x_L$ variations, corresponding to setting $x_L=2$ and $x_L=1/2$. We hence observe at best
a small impact of the matching to higher orders on the total uncertainty, though in some
cases a reduction is visible. In general, the uncertainties are moderate in size. Exceptionally,
they are very large for the first bin including $\lambda = 0$, and the first few bins, towards
smaller values from the peak position, for the Les Houches angularity $\lambda_{1/2}$.
The main effect of the $x_L$ variation is a shift of the distribution towards smaller/larger
observable values. Hence, the bins in the vicinity of the distributions peak are most affected
by these shifts. Given that the angularity distributions typically fall off very
sharply to the left of the maximum, this results in quite significant uncertainties in
this region.


\section{Particle-level predictions and non-perturbative corrections}
\label{sec:np_effects}

To derive actual particle-level predictions, {\em i.e.}\ for final states of detectable hadrons,
we perform corresponding Monte Carlo simulations with \sherpa
and \pythia~\cite{Bierlich:2022pfr}, thereby accounting for non-perturbative effects such
as hadronization and the underlying event (UE) contribution~\cite{Buckley:2011ms,Campbell:2022qmc}.

As demonstrated in a variety of works~\cite{Manohar:1994kq, Dokshitzer:1995zt, Akhoury:1995sp, Korchemsky:1995zm, Dokshitzer:1998qp},
non-perturbative corrections to \mbox{IRC-safe} jet-substructure observables have a power-suppressed form which allows one to consider them as sub-leading corrections for large and moderate jet-$p_T$ values, see, for example, Ref.~\cite{Reichelt:2021svh}.
Nevertheless, as the jet's transverse momentum decreases, one would anticipate that non-perturbative corrections
leave a more pronounced impact on the jet substructure. Moreover, given that collision energies at RHIC are
much smaller compared to the LHC, possibly re-tuning of the parameters of UE and hadronization models might be needed,
to ensure that different physics aspects relevant at LHC and RHIC are properly taken into account. 

Given there is no unique procedure how to incorporate non-perturbative effects into resummed calculations, special
attention needs to be paid to the choice of the correction scheme when comparing first-principles theoretical
predictions against jet-substructure measurements of low-$p_T$ jets. We here correct our resummed predictions for
hadronization and the underlying event using parton-to-hadron level transfer matrices derived from dedicated
\sherpa simulations as introduced in Ref.~\cite{Reichelt:2021svh}.

\subsection*{Hadron-level simulations with \sherpa and \pythia}

For \sherpa we use version 3.0.0$\beta$, considering inclusive dijet production based on its
implementation of the MC@NLO formalism~\cite{Hoeche:2012yf}, dubbed \sherpa MC@NLO in what follows.
The NLO QCD matrix elements for two-jet production get matched with the \sherpa Catani--Seymour dipole
shower~\cite{Schumann:2007mg}. The involved QCD one-loop amplitudes we obtain from the \OpenLoops
library which uses the \collier package~\cite{Denner:2016kdg} for the evaluation of tensor and scalar
integrals.

As for the resummed calculation we employ the NNLO PDF4LHC21\_40\_pdfas set and $\alpha_s(M_Z) = 0.118$.
The central values for the perturbative scales are set to  
\begin{eqnarray}
  \mu_{\text{F}}=\mu_{\text{R}}=\mu_{\text{Q}}=H_T/2\,.\label{eq:scalesjj_MC}
\end{eqnarray}

Scale uncertainties get estimated through $7$-point variations of $\mu_{\text{F}}$
and $\mu_{\text{R}}$, see Eq.~\eqref{eq:7pointvar}, both in the matrix elements and
the parton shower, while we keep the parton-shower starting scale ($\mu_{\text{Q}}$)
fixed~\cite{Bothmann:2016nao}.
The \sherpa UE simulation~\cite{Gleisberg:2008ta} is based on an implementation of
the Sj\"ostrand--Zijl multiple-parton interaction (MPI) model~\cite{Sjostrand:1987su}.
To account for the parton-to-hadron transition we use the \sherpa
cluster fragmentation~\cite{Chahal:2022rid}. Here we include an estimate of the
hadronization parameter uncertainty, based on replica tunes, first presented
in~\cite{Knobbe:2023ehi,Knobbe:2023njd}. To this end, besides the nominal default tune,
we run \sherpa with the hadronization parameters set according to 7 replica tunes,
thereby using the central choice for $\mu_\text{R}$ and $\mu_\text{F}$. Each of the runs
produces an alternative distribution that we treat on an equal footing with the scale
variations and use the envelope of all variations as an estimate for the total uncertainty.

\begin{figure}
  \centering
  \includegraphics[width=1.0\linewidth]{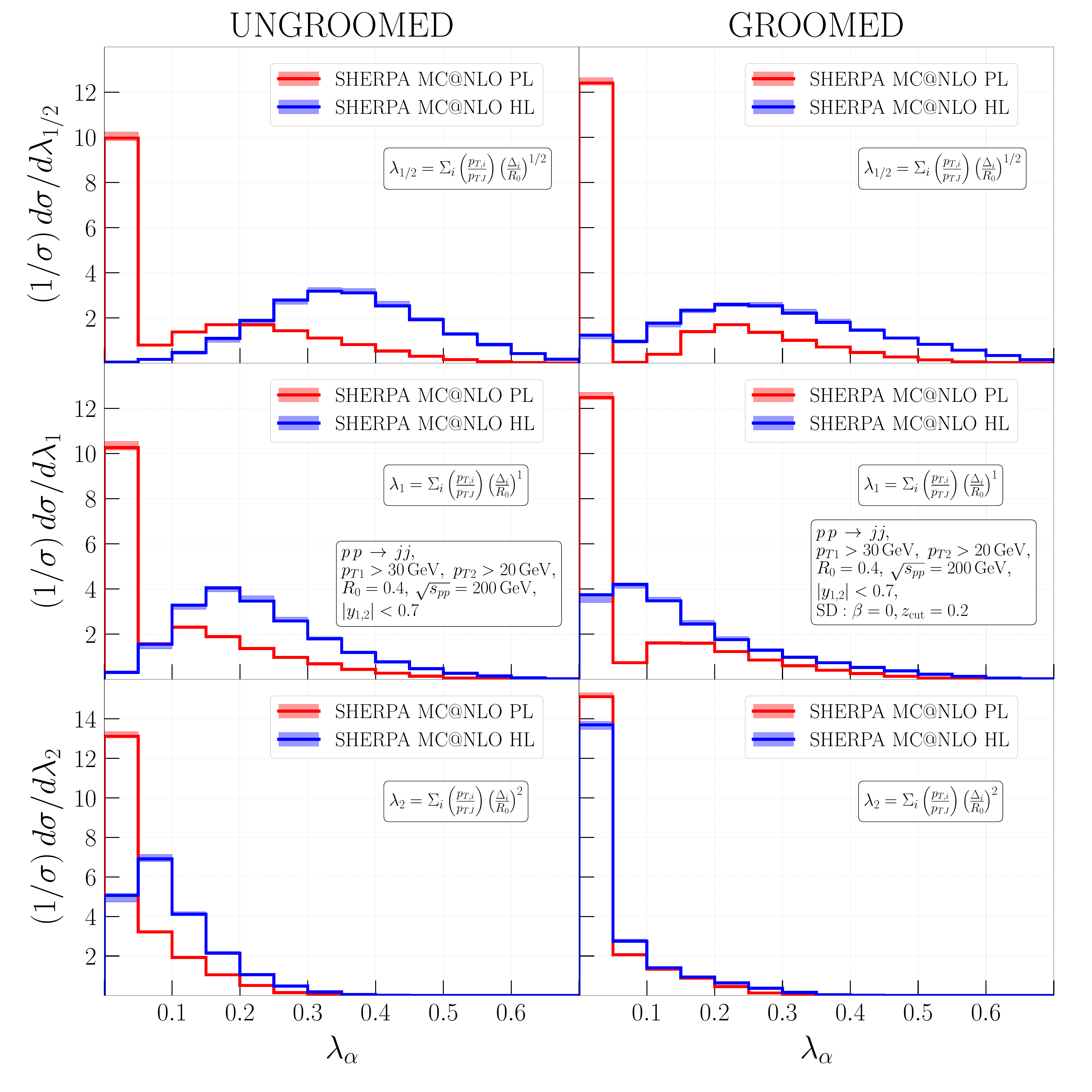}
  \centering
  \caption{Normalized parton- and hadron-level Monte Carlo predictions obtained with
    \sherpa for the $\lambda_{1/2}$, $\lambda_{1}$, $\lambda_{2}$ angularity distributions
    (top to bottom) without and with \softdrop grooming (left and right).}
  \label{fig:ang_all_mc}
\end{figure}

In Fig.~\ref{fig:ang_all_mc} we present the predictions from \sherpa for simulations
at parton level, \emph{i.e.} with hadronization and UE disabled, and at hadron level,
where both effects are taken into account. The general features of the distributions
parallel the observations for the resummed results in Fig.~\ref{fig:ang_all_pl}. We
note that the parton-level distributions show a significant peak in the first bin,
caused by the finite parton-shower cutoff, that effectively prevents particle
production below a certain hardness scale of ${\cal{O}}(1~\text{GeV})$. This leads
to an overpopulation of the first bin from jets with no resolvable emission above
this cutoff. Within the factorized MC approach, the hadronization model is responsible
for re-distributing these jets towards non-vanishing angularities. On the other hand,
the resummed predictions shown earlier do not feature such cutoff, as all integrals
are formally taken across the Landau pole. Hence, it is evident that a simple ratio
of the bin entries at hadron and parton level, at least close
to the shower cutoff, does not suffice to correct resummed calculations and we
therefore have to resort to more sophisticated approaches. We here make use of
a multi-differential transfer-matrix technique introduced in~\cite{Reichelt:2021svh},
detailed below.

Even when ignoring shower-cutoff effects, the impact of hadronization and UE on the
angularity distributions is quite significant, resulting in a change of the peak position
for most cases. In particular the Les Houches angularity $\lambda_{1/2}$ is known to be
very sensitive to non-perturbative effects, see, \textit{e.g.} Refs.~\cite{Caletti:2021oor, Reichelt:2021svh}.
A notable exception appears to be the groomed $\alpha=2$ case, where parton- and hadron-level
predictions from \sherpa seem to largely agree. However, as explained above, for $\alpha=2$
most of the jets with deeply soft radiation end up in the first bin, such that we do not
resolve them very well for $\lambda_2$. In general, \softdrop grooming significantly
ameliorates the impact of non-perturbative effects on the angularity distributions.

To obtain alternative hadron-level predictions we ran further simulations with \pythia
using version 8.310~\cite{Bierlich:2022pfr}. 
To this end, we  simulate inclusive dijet production at the leading order using the \pythia8 default parton shower~\cite{Sjostrand:2004ef, Corke:2010yf}. 
Given the drastic difference of proton--proton collision energy between RHIC
and LHC might necessitate adjustments of the non-perturbative model parameters, we here
employ a dedicated tune of \pythia optimized for the description of RHIC data, referred
to as \emph{Detroit tune}~\cite{Aguilar:2021sfa}. 
For reference, we also derived predictions
based on the \pythia8 default, the so-called \emph{Monash tune}~\cite{Skands:2014pea}.

\begin{figure}
  \centering
  \includegraphics[width=1.0\linewidth]{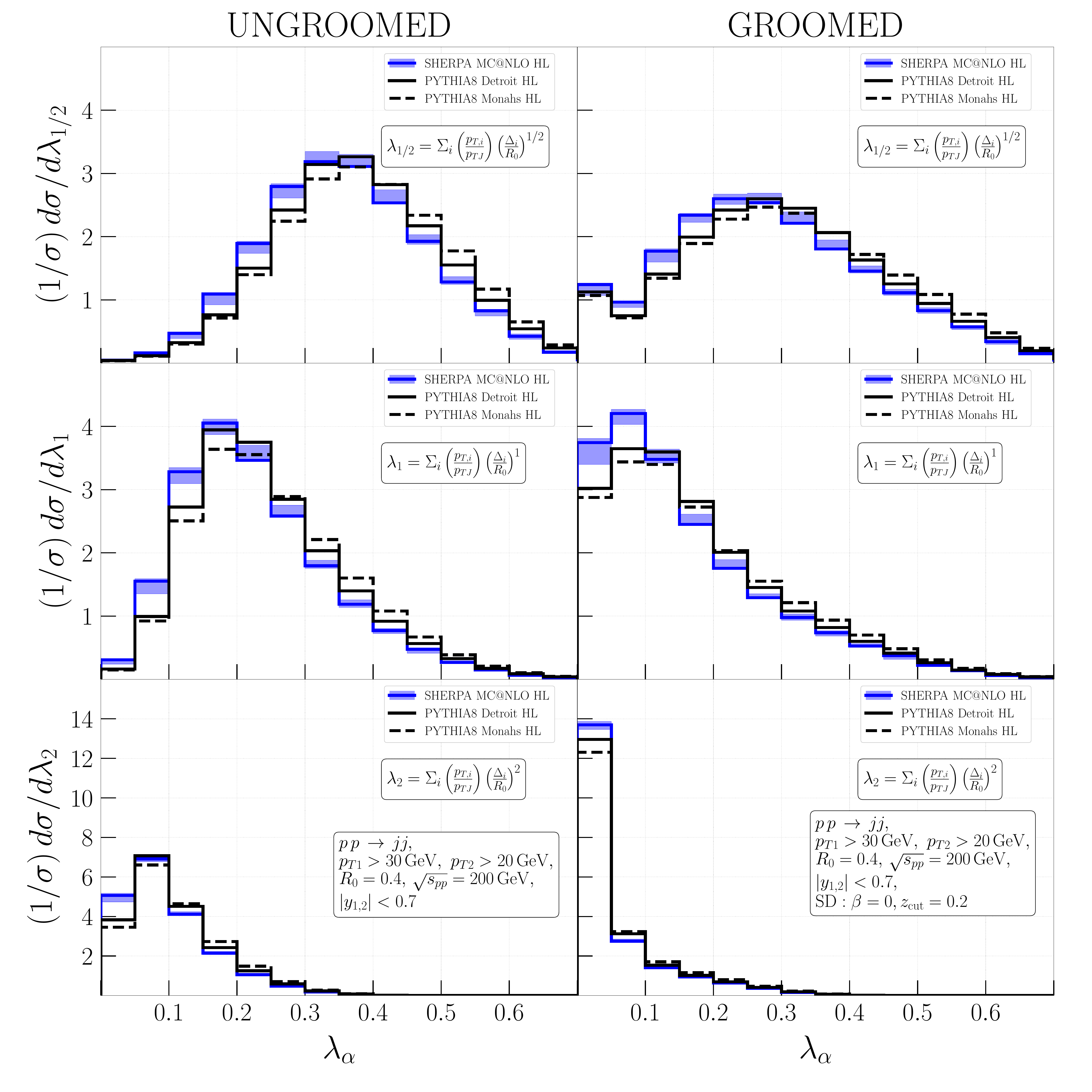}
  \centering
  \caption{Comparison of hadron-level predictions for the ungroomed and groomed angularity observables
    $\lambda_{1/2}$, $\lambda_{1}$ and $\lambda_{2}$ from \sherpa MC@NLO (default tune) and
    \pythia8 results obtained for the ``Detroit'' tune~\cite{Aguilar:2021sfa} and the default Monash
    2013 tune~\cite{Skands:2014pea}.}
\label{fig:MC_tuning}
\end{figure}

\begin{figure}
  \centering
  \includegraphics[width=0.5\linewidth]{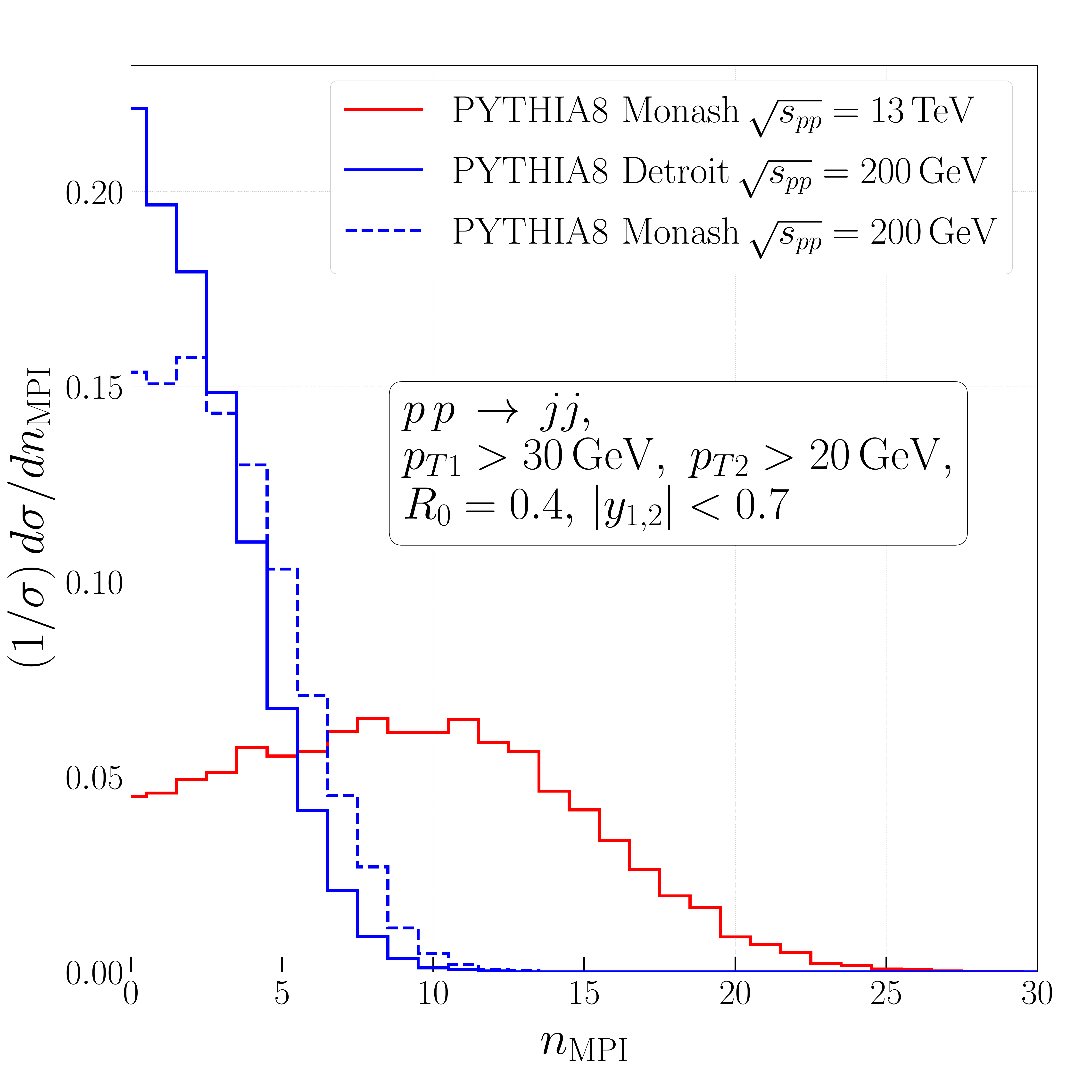}
  \centering
  \caption{Number of secondary scatterings per proton--proton collision at RHIC (blue lines) and LHC (red line) as
    predicted by the \pythia8 event generator. For the LHC case we show predictions based on the default Monash tune.
    For the RHIC scenario we present results both for the Detroit (solid) and the Monash (dashed) tune. }
\label{fig:MPI_multiplicity}
\end{figure}

In Fig.~\ref{fig:MC_tuning} we present a comparison of hadron-level predictions obtained with \pythia
based on the Detroit and Monash tunes, as well as those from \sherpa using its default tune. 
In general, the particle-level results from the two generators agree quite well regarding the shapes
of the plain and the soft-drop groomed angularity distributions. However, the \pythia results
for both tunes are slightly shifted towards larger angularity values, corresponding to a broader
jet-constituent distribution. This effect is somewhat more pronounced for the Monash tune, \pythia's
default. The dedicated Detroit tune appears to produce results in-between the \sherpa MC@NLO and
the \pythia Monash-tune predictions.  
The dominant difference between the two \pythia tunes consists in the parameter settings affecting the
UE activity, see Ref.~\cite{Aguilar:2021sfa} for details. Given the significantly lower proton-beam
energy at RHIC in comparison to the LHC, the average number of secondary scatterings per proton--proton
collision at RHIC is much smaller. This is illustrated in Fig.~\ref{fig:MPI_multiplicity}, 
where we show the distribution of MPI scatters associated with a dijet-production hard process, using
the event selection criteria listed in Sec.~\ref{sec:def}, for $\sqrt{s_{\rm pp}}=200\,\text{GeV}$ and
$\sqrt{s_{\rm pp}}=13\,\text{TeV}$, respectively. For the LHC setup we show predictions based on the Monash
tune, the \pythia8 default in particular for LHC simulations. The distribution is rather broad and peaks
around $n_\text{MPI}\approx 10$.
In contrast, at RHIC energies the distribution based on the Monash-tune parameters has its maximum
at $n_\text{MPI} = 2$ and is rather narrow. Considering the Detroit tune, the distribution peaks at
$n_\text{MPI} = 0$ and is quickly falling. Accordingly, the majority of events has either no or
only 1-2 associated secondary scatterings. These differences in MPI activity for the two tunes
explain the shifts in the angularity distributions observed in Fig.~\ref{fig:MC_tuning}. In conclusion,
for the here considered event-selection criteria most of the hadronic collision energy at RHIC is
deposited in the hard scattering, while the impact of the UE is significantly reduced compared to
the LHC. As a consequence, non-perturbative corrections that affect jet-substructure observables at
RHIC are largely dominated by hadronization effects. Accordingly, studies of jet angularities sensitive
to non-perturbative corrections, as in particular $\lambda_{1/2}$, provide stringent tests for hadronization
models as implemented in general-purpose MC generators.


\subsection*{The transfer-matrix approach to non-perturbative corrections}\label{sec:transfermatrix}
To account for the alteration of the derived resummed jet angularity distributions
due to non-perturbative corrections we employ parton-to-hadron level transfer matrices
as introduced in Ref.~\cite{Reichelt:2021svh}.  This quite general approach is applicable
to an arbitrary set of observables, measured in a multi-differential way. In
general, it aims to keep track both of the migration in event kinematics, used
to define the fiducial phase space, that get partially integrated over, as well
as changes in the actual observable of interest, \emph{i.e.}\ in the here
considered case the angularity variable.

We consider a generic scattering process which results in a partonic configuration $\mathcal{P}$.
Through NP effects the set of parton momenta gets mapped onto a hadron-level configuration
$\mathcal{H}\left(\mathcal{P} \right)$. The map $\mathcal{H}$ thereby accounts for hadronization
and UE corrections. It could be derived from field-theoretical considerations (see for
example Refs.~\cite{Hoang:2019ceu,Pathak:2020iue} for recent work on
\softdrop observables) or it can be extracted from parton-shower simulations interfaced
to a model of NP phenomena.
For a given configuration $\mathcal{P}$ or $\mathcal{H}\left(\mathcal{P} \right)$, we then
measure a set of $m$ observables, $\vec{V}\left(\mathcal{P}\right)$ or
$\vec{V}\left(\mathcal{H}\left(\mathcal{P} \right)\right)$.\footnote{For simplicity we here
have chosen the same set of observables $\vec{V}$ on the parton- and hadron-level configurations which,
however, is not strictly necessary.}
We define the transfer operator as the conditional probability to
measure a set of observables $\vec{v}_h$ at hadron level on
$\mathcal{H}\left(\mathcal{P} \right)$, given that the parton-level
observables were $\vec{v}_p$:
\begin{equation} \label{transfer-operator}
  \mathcal{T}(\vec{v}_h|\vec{v}_p) = \frac{\int \mathop{d\mathcal{P}}
    \frac{\mathop{d\sigma}}{\mathop{d \mathcal{P}}}
    \delta^{(m)}\left(\vec{v}_p-\vec{V}\left(\mathcal{P}\right)\right) \delta^{(n)} \left(\vec{v}_h-\vec{V}\left(\mathcal{H}\left(\mathcal{P}\right)\right)\right)}
  {\int \mathop{d\mathcal{P}}
    \frac{\mathop{d\sigma}}{\mathop{d \mathcal{P}}}
    \delta^{(m)}\left(\vec{v}_p-\vec{V}\left(\mathcal{P}\right)\right)}\,.
\end{equation}
This way, the multi-differential distribution for the set of hadron-level observables $\vec{v}_h$ can be written as
\begin{equation}
  \frac{\mathop{d^m \sigma^\text{HL}} }{\mathop{dv_{h,1} \dots dv_{h,m}}} = \int \mathop{d^m \vec{v}_p}\,  \mathcal{T}(\vec{v}_h|\vec{v}_p) \, \frac{\mathop{d^m\sigma^\text{PL} }}{\mathop{dv_{p,1} \dots dv_{p,m}}}\,.
\end{equation}
When working with binned distributions, one obtains binned cross sections by integrating
the multi-differential distribution over hypercubes in the observables' space. 
At parton level the cross section in any given hyper-bin $p$ can be written as 
\begin{align}
   \Delta \sigma^\text{PL}_{p} &= \int \mathop{d\mathcal{P}}
  \frac{\mathop{d\sigma}}{\mathop{d \mathcal{P}}}
  \Theta_{p}\left(\mathcal{P}\right)\,,
\end{align}
where
\begin{align} \label{thetap}
  \Theta_{p}\left(\mathcal{P}\right)=
    \prod_{i=1}^m \theta(V_i(\mathcal{P})-v^\text{min}_{p,i})\theta(v^\text{max}_{p,i}-V_i(\mathcal{P}))\,.
\end{align}
When considering a binned distribution at hadron level, the transfer
operator from parton-level bin $p$ to a given hadron-level bin $h$
becomes a matrix of the form
\begin{equation}
  \mathcal{T}_{hp} = \frac{\int \mathop{d\mathcal{P}}
    \frac{\mathop{d\sigma}}{\mathop{d \mathcal{P}}}
    \Theta_p\left(\mathcal{P}\right)\Theta_h\left(\mathcal{H}\left(\mathcal{P}\right)\right)
  }
  {\int \mathop{d\mathcal{P}}
    \frac{\mathop{d\sigma}}{\mathop{d \mathcal{P}}}
    \Theta_p\left(\mathcal{P}\right)
  }\,,
\end{equation}
with
\begin{align} \label{thetah}
  \Theta_{h}\left(\mathcal{\mathcal{H}(\mathcal{P})}\right)=
   \prod_{i=1}^m \theta \left( V_i\left(\mathcal{\mathcal{H}(\mathcal{P})}\right)-v^\text{min}_{h,i}\right) \theta \left(v^\text{max}_{h,i}-V_i\left(\mathcal{\mathcal{H}(\mathcal{P})}\right)\right)\,.
\end{align}
Consequently, the final hadron-level distribution in the hyper-bin $h$
is obtained by the weighted sum of all parton-level contributions
\begin{equation}\label{eq:Tfinal}
  \mathop{\Delta\sigma_h^\text{HL}} = \sum_{p} \mathcal{T}_{hp} \mathop{\Delta\sigma_p^\text{PL}}\,.
\end{equation}

The elements of the transfer matrices appearing in
Eq.~\eqref{eq:Tfinal} can easily be extracted from a multi-purpose
generator in a single run, provided that individual events are
accessible at different stages of their evolution in the simulation process.
Note however that, while parton shower and hadronization are
treated in a factorized form in all multi-purpose event
generators~\cite{Buckley:2011ms}, this is not necessarily the case
for the UE.
In particular, \mbox{\pythia 8}~\cite{Sjostrand:2014zea} makes
use of an interleaved evolution of the initial-state shower and the
secondary interactions~\cite{Sjostrand:2004ef, Corke:2010yf}. Accordingly,
in a full event simulation within such model there is no
notion of an intermediate parton-level final state that is directly
comparable to a resummed calculation.
However, in \sherpa the parton
showers off the hard process and the simulation of multiple-parton
interactions are fully separated, \emph{i.e.}\ the UE is simulated
only after the shower evolution of the hard interaction is
completed. The secondary scatterings then get showered and ultimately the
partonic final state consisting of the showered hard process and
multiple-parton interactions gets hadronized.

Accordingly, we here derive parton-to-hadron level transfer matrices
for the jet angularities using the \sherpa generator. We obtain them from
the same runs as the histograms shown as MC@NLO hadron-level \sherpa predictions
in what follows. Besides the default hadronization tune we further determine
a separate transfer matrix for each of the 7 replica
tunes~\cite{Knobbe:2023ehi,Knobbe:2023njd} for the \sherpa cluster hadronization
model~\cite{Chahal:2022rid}. We fold the resummed result, including each of the
scale and $x_L$-parameter variations performed, with each of these matrices. The
resulting envelope is then used as estimate for the total uncertainty of the
prediction at hadron level.

When extracting the transfer matrices from \sherpa, we do allow for the possibility that
non-perturbative corrections can reduce the transverse momentum of a given jet,
and hence our cross section that is defined by requiring two jets of a certain
hardness may become smaller. The sum of the probabilities for one parton-level
bin to end up in any
of the hadron-level bins is hence not guaranteed to be 1, and we indeed observe
a significant loss of cross section when going from parton to hadron level. We
however neglect the reverse effect, \emph{i.e.} that a rather soft jet at parton
level could pass the selection cut at hadron level, for example by picking up
contributions from the underlying event. This is an important effect for example
at the LHC, but appears not to be significant for our RHIC setup, consistent
with the much reduced underlying event activity observed earlier.

\begin{figure}
  \centering
  \includegraphics[width=0.47\linewidth]{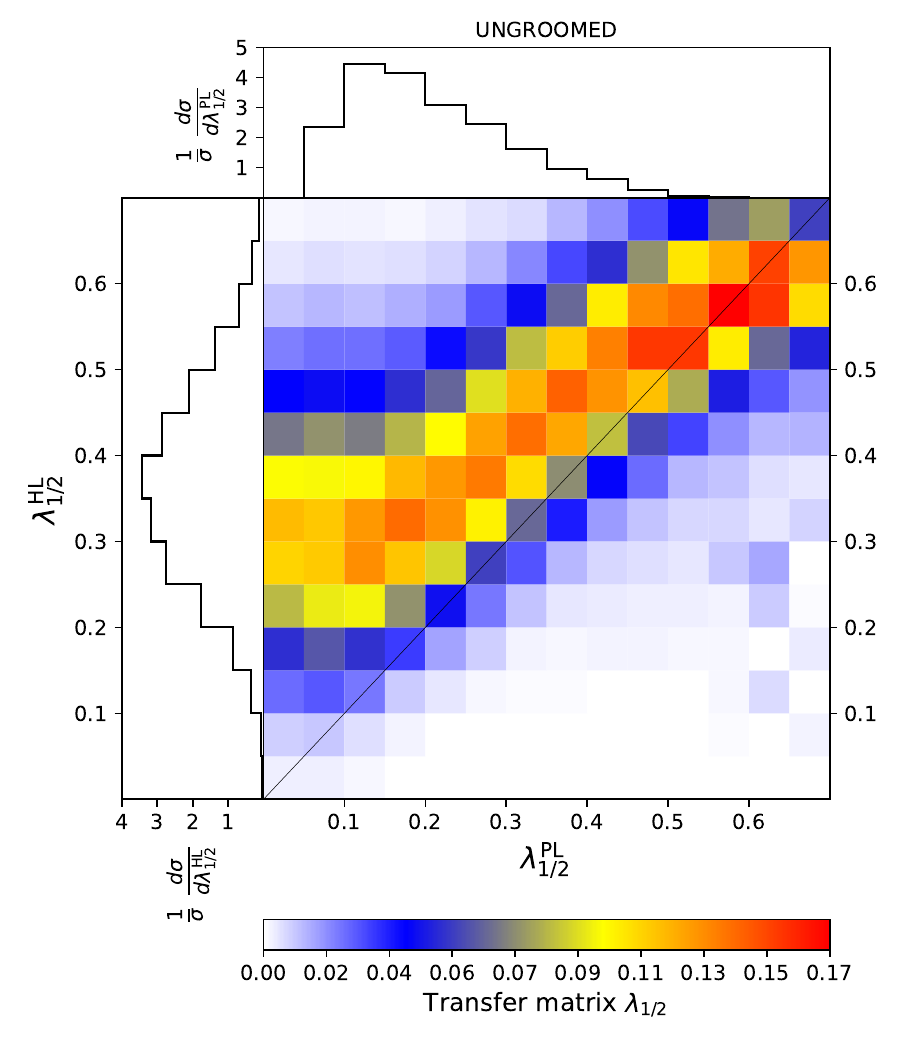} 
  \hspace{1em}
  \includegraphics[width=0.47\linewidth]{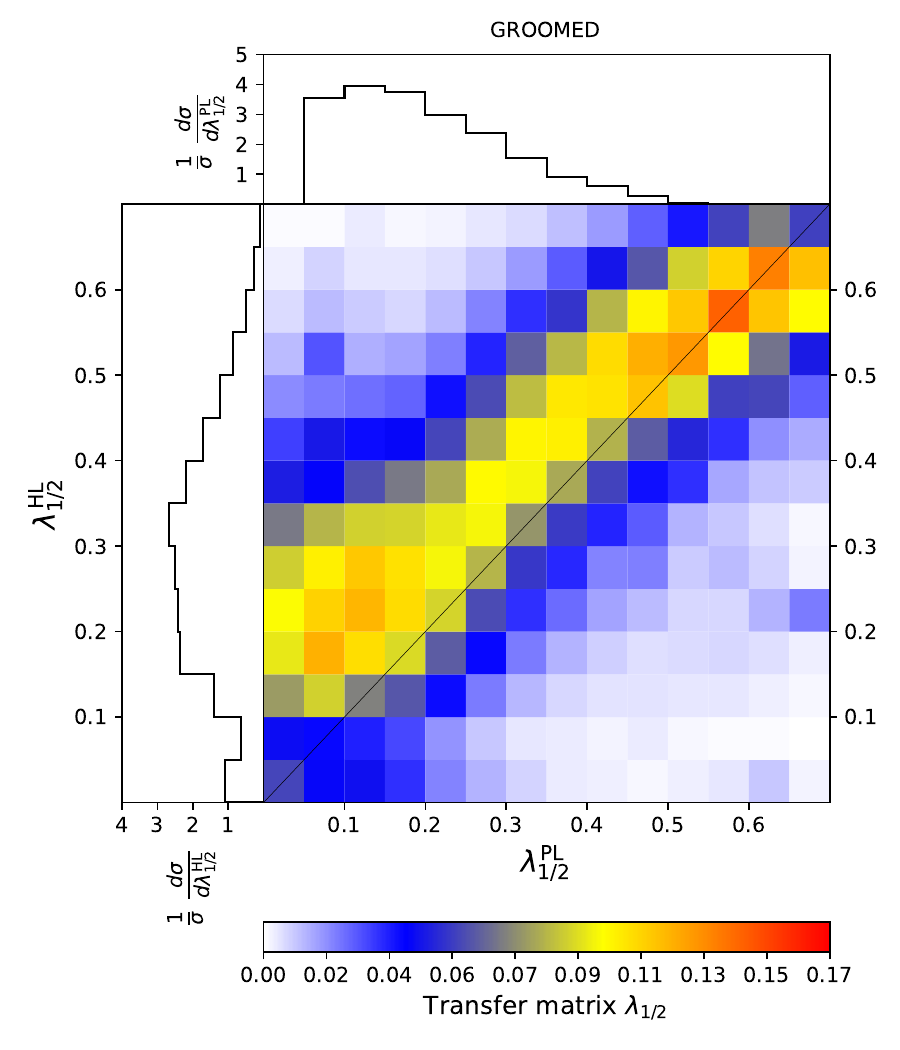}
  \centering
  \caption{Transfer matrices defined according to Eq.~\eqref{eq:Tfinal} for
    the ungroomed (left) and groomed (right) angularity $\lambda_{1/2}$, extracted from
    \sherpa MC@NLO hadron-level simulations for the default hadronization tune. }
\label{fig:TM_Sherpa}
\end{figure}

To give an illustrative example we present in Fig.~\ref{fig:TM_Sherpa} the binned
transfer matrices defined via Eq.~\eqref{eq:Tfinal} for the plain and \softdrop-groomed
LHA $\lambda_{1/2}$. In particular we here show the results corresponding to the nominal
\sherpa hadronization tune and the event selections detailed in Sec.~\ref{sec:def}.
In the 2D presentations the abscissa corresponds to the observable value at parton level,
\emph{i.e.} $\lambda^{\text{PL}}_{1/2}$, while the ordinate represents the observable at
hadron level, \emph{i.e.} $\lambda^{\text{HL}}_{1/2}$. Hence, a vertical line represents
the binned probability for an event with given $\lambda^{\text{PL}}_{1/2}=v^{\text{PL}}$ to
end up at some $\lambda^{\text{HL}}_{1/2}$ along the line. Similarly, a horizontal line
at $\lambda^{\text{HL}}_{1/2}=v^{\text{HL}}$ indicates the likeliness for this hadron-level
observation to originate from a given parton-level value $\lambda^{\text{PL}}_{1/2}$.
Along with the transfer matrices, we here show the corresponding \NLOpNLLp distribution
at parton level (top of the panels) and hadron level (left-hand side of the panels), respectively.
The latter are thereby obtained by convolving the former with the given transfer matrix,
resulting in our final \NLOpNLLp+NP prediction. 

For the ungroomed case we observe a significant shift of the parton-level observable
towards higher values. In fact the transfer matrix has almost no entries in the lower
triangle. Furthermore, the ridge of the 2D distribution appears rather broad and
parallel shifted with respect to the diagonal. In particular for events with
$\lambda^{\text{PL}}_{1/2}\lsim 0.1$ the shift is even larger. Only for the tails of
the distributions, \emph{i.e.}\ $\lambda_{1/2}\geq 0.5$ is the transfer matrix largely
diagonal and the hadron-level distribution receives non-negligible contributions also
from higher parton-level values. For the groomed case, shown in the right panel, the
hadronic corrections are on average somewhat reduced. There now appear a few entries also below
the diagonal, corresponding to a reduction of the observable value from parton to
hadron level. However, most of the time events still get shifted towards larger
$\lambda^{\text{HL}}_{1/2}$, though the ridge of the 2D distribution now lies closer
to the diagonal. As for the ungroomed case, beyond $\lambda_{1/2}\geq 0.5$ the transfer
matrix is largely symmetric.

To capture the observed highly non-trivial migration in the angularity observables
from parton to hadron level indeed requires the use of a somewhat sophisticated
correction method, such as the transfer matrices employed here. In principle,
observable-specific analytical corrections can be derived, modelling an additional
non-perturbative emission, that is supposed to be soft~\cite{Dasgupta:2007wa}.
The effect of such emission can be accounted for by an observable shift according to
\begin{equation}
  \lambda^{\text{HL}}_\alpha=\lambda^{\text{PL}}_\alpha+\delta\lambda^{\text{NP}}_\alpha(\Omega)\,,
\end{equation}
where $\delta\lambda^{\text{NP}}_\alpha$ will depend on the observable value $\lambda^{\text{PL}}_\alpha$
and one or several non-perturbative parameters $\Omega$, that need to be estimated.
An application to \softdrop thrust in electron--positron annihilation has been presented
in  Ref.~\cite{Marzani:2019evv}. 

\begin{figure}
  \centering
  \includegraphics[width=1.0\linewidth]{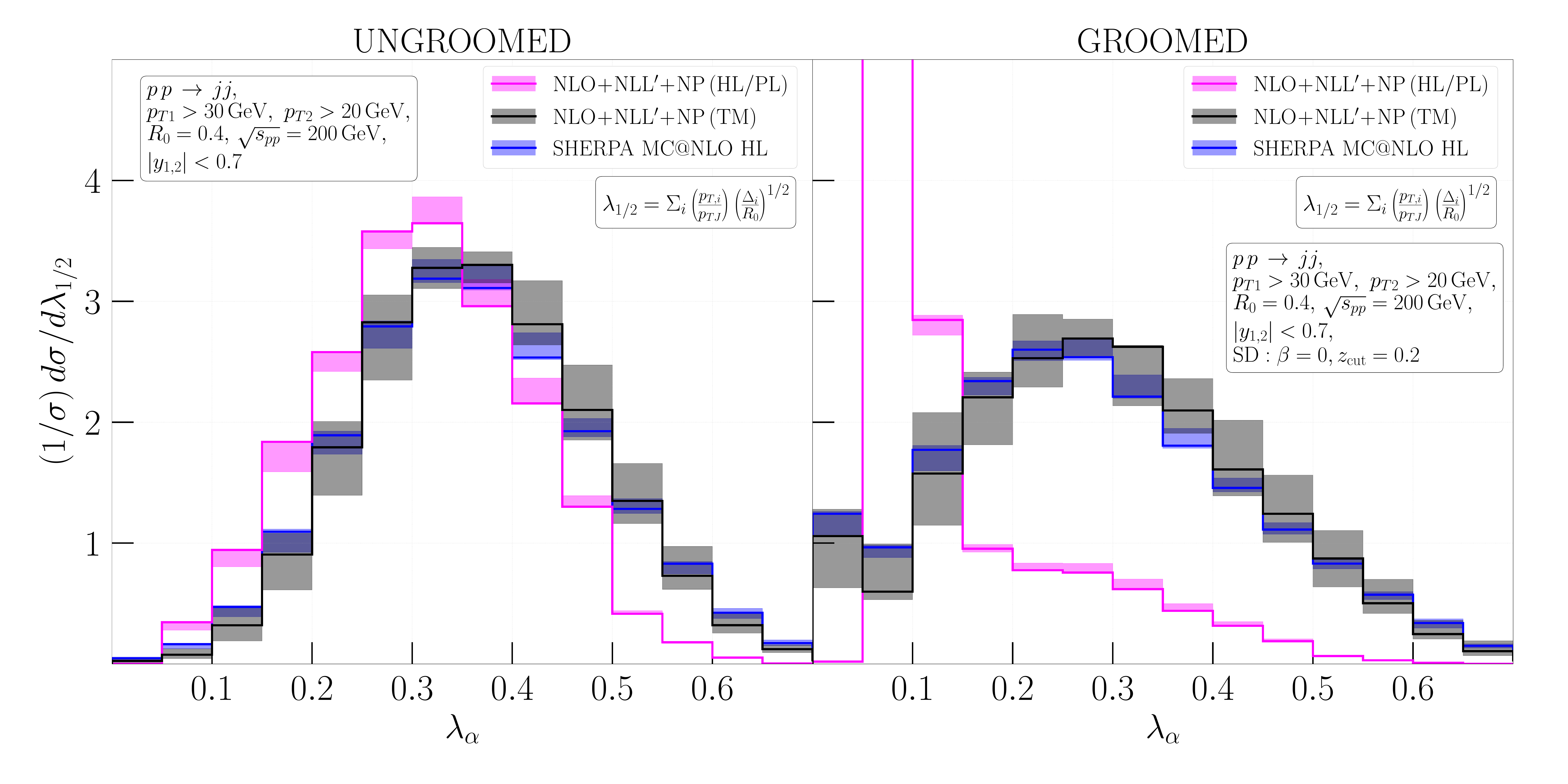}
  \centering
  \caption{Hadron-level distributions for the $\lambda_{1/2}$ angularities without (left)
    and with \softdrop grooming (right). Shown are hadron-level predictions from \sherpa
    along with two variants for correcting the \NLOpNLLp parton-level result for non-perturbative
    corrections, through the transfer-matrix approach (grey) and the binwise multiplicative
    correction (purple).}
  \label{fig:ang_all_lha_ratio}
\end{figure}

Another alternative, based on Monte Carlo simulations at hadron and parton level, is a simple
binwise multiplicative correction, according to
\begin{equation}\label{eq:NPratioCorr}
  \lambda^\text{HL}_\alpha = \lambda^\text{PL}_\alpha \times\left(\frac{\lambda^{\text{HL,MC}}_{\alpha}}{\lambda^{\text{PL,MC}}_{\alpha}}\right)\,.
\end{equation}
To illustrate the latter method we present in Fig.~\ref{fig:ang_all_lha_ratio} corresponding
predictions for the above considered plain and groomed $\lambda^{\text{HL}}_{1/2}$ observables.
Besides the hadron-level predictions from \sherpa and the \NLOpNLLp+NP predictions obtained
through the transfer-matrix approach, we show results for the multiplicative correction
scheme. It is apparent that the binwise correction of the resummed distribution according to
Eq.~\eqref{eq:NPratioCorr} yields significantly different results than the transfer-matrix
method. The transfer-matrix corrections yield results very compatible with the \sherpa
particle-level simulation. However, for the multiplicative correction, when considering
the ungroomed case, the hadron-level result is significantly shifted towards lower
angularity values. For the groomed case, in particular in the region below and around the
parton-shower cutoff, \emph{cf.} Fig.~\ref{fig:ang_all_mc}, the correction weights drastically
overestimate the impact of non-perturbative effects. In consequence, for the normalized
distribution the tail appears harshly suppressed. Accordingly, in what follows we will
employ the transfer-matrix method to incorporate non-perturbative effects into our resummed
predictions.

In Appendix~\ref{app:shapefuncs} we provide an additional brief discussion regarding the
relation of our transfer matrices and the shape-function approach of \cite{Korchemsky:1999kt}.
Therein, we also show illustrations of the transfer matrices for the groomed and ungroomed
jet-width and jet-thrust angularities.

\FloatBarrier
\subsection*{Hadron-level predictions for jet angularities in $\rm pp$ collisions}

Having derived parton-to-hadron level transfer matrices specific for the six
variants of jet angularities, \emph{i.e.} $\lambda_{1/2}$, $\lambda_{1}$ and $\lambda_{2}$
with and without \softdrop grooming, in dijet production in $\rm pp$ collisions at RHIC,
with the fiducial phase space defined in Sec.~\ref{sec:def}, we can employ those
to correct our \NLOpNLLp predictions for non-perturbative effects.

\begin{figure}
  \centering
  \includegraphics[width=1.0\linewidth]{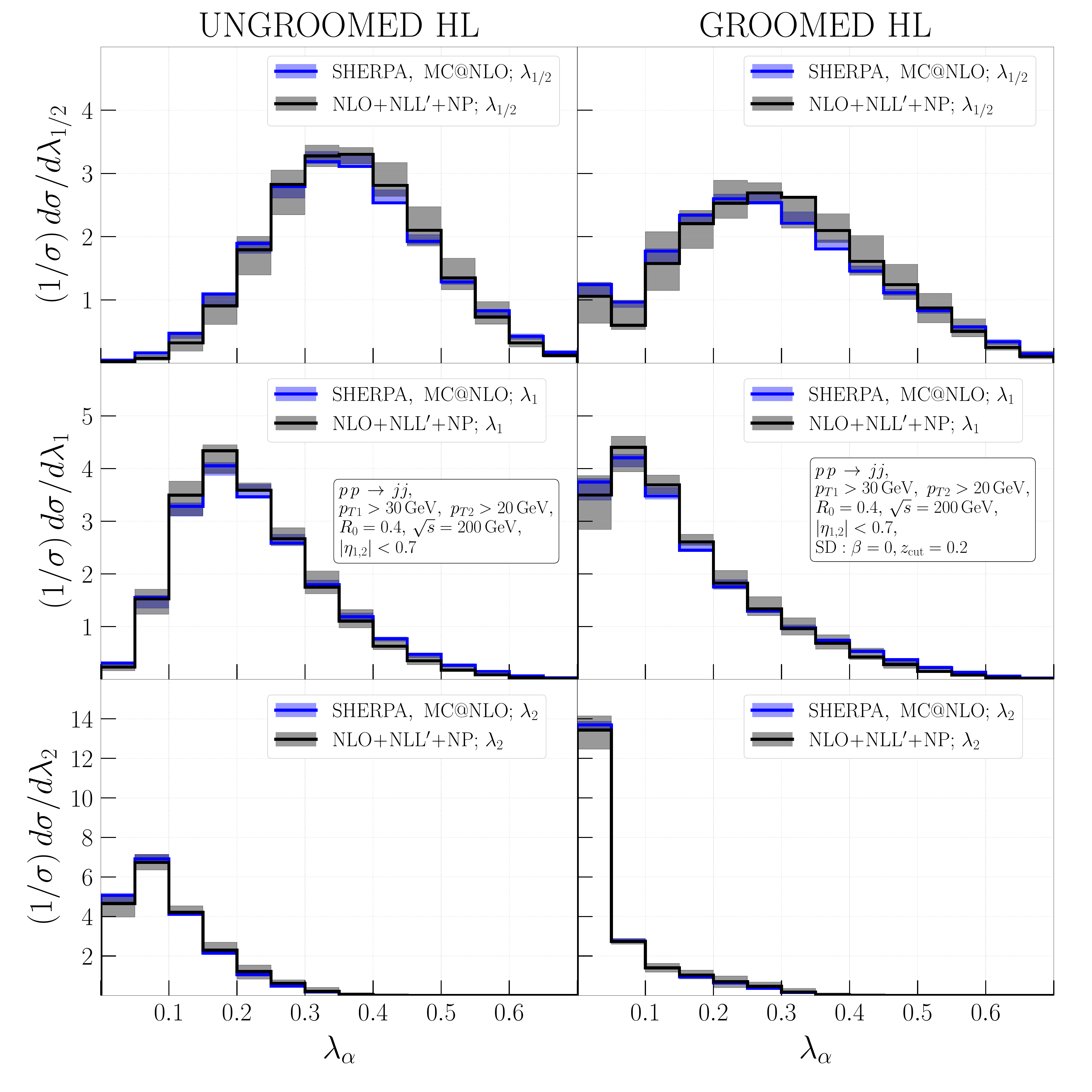}
  \centering
  \caption{Hadron-level distributions for the $\lambda_{1/2}$, $\lambda_{1}$, $\lambda_{2}$
    angularities without (left) and with \softdrop grooming (right). The shown \NLOpNLLp
    predictions are corrected for non-perturbative effects using the transfer-matrix approach,
    dubbed \NLOpNLLp+NP, and compared to corresponding particle-level simulations with \sherpa of MC@NLO
    accuracy.}
  \label{fig:ang_all_hl}
\end{figure}

In Fig.~\ref{fig:ang_all_hl} we present our final hadron-level predictions referred
to as \NLOpNLLp+NP, based on the transfer matrices derived from \sherpa simulations.
Alongside we show the corresponding particle-level predictions from \sherpa, based on
parton-shower simulations at MC@NLO accuracy, supplemented with an UE simulation and
hadronization. Such, these two predictions could in principle directly be compared to
data from the sPHENIX experiment at RHIC when those are corrected for detector effects.

For both cases, we construct variants of the predictions from 7-point scale variations
of the renormalization and factorization scale through on-the-fly reweighting~\cite{Bothmann:2016nao},
as well as from variations of hadronization-model parameters through 7 replica
tunes~\cite{Knobbe:2023ehi,Knobbe:2023njd}. Note, for the \sherpa MC@NLO simulation,
variations of the scales also affect the scales used in the initial- and final-state
shower evolution. Within the resummed calculation however, we can vary the $x_L$ parameter
to obtain a more solid estimate of the logarithmic uncertainty. In both cases, the final
uncertainty again is taken as the envelope of all variants, respectively.

For the normalized angularity distributions shown in Fig.~\ref{fig:ang_all_hl} we
observe excellent agreement between the two predictions. In fact, both results are fully
compatible within their estimated uncertainties. In terms of the central values,
the \NLOpNLLp+NP calculation appears to predict a slightly softer, \emph{i.e.} with a
peak shifted to smaller angularity values, distribution for the Les Houches
angularity $\lambda_{1/2}$. The distribution is also slightly wider, such that
the tail towards the kinematic endpoint at large $\lambda_{1/2}$ is somewhat higher
than in the \sherpa prediction. The general trend is reversed in the $\lambda_1$
case, where the Monte Carlo curve appears marginally shifted to smaller
values. For $\lambda_2$, the two distributions are very close and we can not
make out any general trend. Again, any of these effects are covered by the
uncertainties of the \NLOpNLLp+NP calculation, and we conclude that at our
accuracy there is no consistent systematic difference between the results.
For the general physics features of the distributions we refer the reader back
to our discussion in Sec.~\ref{sec:resum} and above in this part.

The estimated uncertainty of the \sherpa MC@NLO predictions is generally smaller
than that of the matched resummed ones, consistent with the observations made
in~\cite{Reichelt:2021svh}. While the uncertainty estimate for the \NLOpNLLp+NP
prediction is somewhat larger, its size appears significantly reduced with
respect to one for the partonic \NLOpNLLp predictions shown in
Fig.~\ref{fig:ang_all_pl}. At first sight this might seem counter intuitive,
however, it can be traced back to the convolution with the parton-to-hadron
level transfer matrices that largely smear out the sharp peaks at parton level.

Looking back at Fig.~\ref{fig:TM_Sherpa} it is apparent that the bulk of the hadron-level
distribution, around $\lambda^{\text{HL}}_{1/2}\sim 0.3$, is largely made up of commensurate
contributions from several bins of lower angularity at parton level, around
$\lambda^{\text{PL}}_{1/2}\sim0.1-0.25$. While in particular the $x_L$ variations widely shift
the peak of the parton-level distribution between these low angularity bins, this does
not significantly alter their combined contribution to the bulk of the hadron-level result.
Furthermore, the bins at very low $\lambda^{\text{PL}}$, featuring the largest uncertainties,
are systematically pushed and broadly smeared out towards larger values of $\lambda^{\text{HL}}$.

Supported also by the good agreement of the \sherpa and \pythia results, presented in
Fig.~\ref{fig:MC_tuning}, we are confident that the here presented \NLOpNLLp+NP predictions
provide reliable theoretical expectations for experimental measurements with sPHENIX at RHIC.
Through a detailed comparison of jet-angularity observables in proton--proton collisions
at RHIC energies the modelling of all-order perturbative corrections as well as
non-perturbative aspects, in particular hadronization, can be thoroughly tested. For the
here employed Monte Carlo simulations such data would furthermore provide useful means
to tune and optimize model parameters associated with hadronization and the underlying
event. Having established a good understanding of jet-substructure observables in
$\rm pp$ collisions, these measurements can serve as a baseline for corresponding analyses
in nucleus--nucleus scattering, thereby trying to infer about possible QCD-medium modifications.

\section{Estimating medium-interaction effects on jet angularities}
\label{sec:medium_effects}

A high-density medium of deconfined quarks and gluons, referred to as Quark-Gluon Plasma
(QGP)~\cite{Csernai:1994xw}, can be formed in high-energy heavy-ion collisions.
The characteristic scale of a QGP is expected to be an order of magnitude
larger than the QCD confinement scale, $\Lambda_{\rm QCD} \approx 0.2 \, {\rm GeV}$.
Hard QCD processes initiating the production of particles with sufficiently high transverse
momenta, such that the hierarchy $p_{T,{\rm jet}} \gg \Lambda_{\rm QGP} \gg \Lambda_{\rm QCD}$ holds,
can be considered as factorized from later interactions with the QGP medium~\cite{Accardi:2004gp}.
However, the multiple soft and collinear subsequent emissions leading to jet formation are expected to be
affected by medium-induced interactions giving rise to a broad range of jet-quenching phenomena~\cite{Bjorken:1982tu,
  Gyulassy:1990ye, Wang:1991hta, Wang:1992qdg, Gyulassy:1993hr,
  Casalderrey-Solana:2007knd, dEnterria:2009xfs}.
Qualitatively, jet quenching embraces three distinct effects: Energy loss, \emph{i.e.}\ the reduction
of the energy of particles inside a jet or the jet energy as a whole~\cite{Bjorken:1982tu}, leading to the
suppression of cross sections; an increase in the number of soft particles inside the jet,
\emph{i.e.}\ the softening of jet fragmentation~\cite{Gyulassy:1990ye, Wang:1992qdg}; and jet broadening,
\emph{i.e.}\ an increase in the number of soft particles displaced away from the jet axis~\cite{Salgado:2003rv, Borghini:2005em, Armesto:2008qh,Armesto:2008qe, PerezRamos:2008swo, PerezRamos:2008uyr}.

In fact, all these jet-quenching effects are expected to affect the jet
angularities as defined in Eq.~\eqref{eq:ang-def}.
Various dedicated MC programs were developed that aim to simulate jet-quenching effects,
including HIJING~\cite{Wang:1991hta}, \jewel~\cite{Zapp:2013vla}, PQM~\cite{Dainese:2004te}, HYBRID~\cite{Casalderrey-Solana:2014bpa} and
\mbox{Q-\pythia}~\cite{Armesto:2009fj}.
Apart from these codes there exists the general JETSCAPE framework~\cite{Putschke:2019yrg} capable
of simulating jet propagation through a QGP medium based on the usage of various hydrodynamical
models~\cite{Schenke:2010nt,Karpenko:2013wva, Shen:2014vra, Bazow:2016yra, Pang:2018zzo}.
In this paper we would like to provide a 
qualitative estimate of
quenching effects for jet angularities in AA collisions at RHIC energies, \emph{i.e.}\
$\sqrt{s_\text{NN}}=200\,\text{GeV}$. Given hydrodynamical simulations can easily be very time-consuming,
we reserve using the JETSCAPE framework for future work and
consider only two rather light-weighted MC choices, namely \mbox{Q-\pythia} and
\jewel, both based upon modified versions of the \pythia6 virtuality-ordered parton
shower~\cite{Sjostrand:2006za}.

\FloatBarrier
\subsection*{Medium-effect simulations based on Q-\pythia and \jewel}

The Q-\pythia model is based upon the addition of an extra term to  the
standard Altarelli--Parisi shower splitting kernels $P_{\rm vac}(z)$~\cite{Altarelli:1977zs},
according to
\begin{eqnarray}
	P_{\rm tot}(z) = P_{\rm vac}(z) + \Delta P(z, t, \hat{q}, L, E)\,,
	\label{eq:medium_splitting_func}
\end{eqnarray}
where $\Delta P(z, t, \hat{q}, L, E)$ accounts for medium modifications in dependence on the
virtuality $t$ of the radiating parton, its energy $E$, the medium length $L$, and the
transport coefficient $\hat{q}$.
Whereas vacuum splitting functions $P_{\rm vac}(z)$ are well known and can be
derived from first-principles perturbative QCD calculations, the additional term in
Eq.~\eqref{eq:medium_splitting_func} is in a priori unknown and in general cannot be
evaluated by means of perturbation theory only.
However, the medium-induced splitting functions can be obtained within the
Baier--Dokshitzer--Mueller--Peign\'{e}--Schiff--Zakharov (BDMPS-Z)
formalism~\cite{Baier:1996kr, Baier:1996sk, Zakharov:1996fv, Zakharov:1997uu,
  Zakharov:1998sv, Wiedemann:2000tf, Wiedemann:2000za}, where interactions between
QCD parton-shower particles and the QGP medium get approximated by an infinite number
of soft scatterings\footnote{For phenomenological applications of the BDMPS-Z
approach see Refs.~\cite{Baier:2000mf, Salgado:2002cd, Gyulassy:2003mc,
  Kovner:2003zj, Salgado:2003gb, Casalderrey-Solana:2007knd, Zapp:2011ya}.}.
The residual dependence on non-perturbative physics is encoded in the
time-dependent density of scattering centers $n(\xi)$ and the single dipole elastic
scattering  cross section $\sigma({\bf{r}})$ which are related to the time-dependent transport coefficient $\hat{q}$ via
\begin{eqnarray}
	n(\xi) \sigma({\bf r}) = \frac{1}{2} \hat{q}(\xi) {\bf r}^2,
\end{eqnarray}
where $\hat{q}(\xi)$ is interpreted as the
medium induced transverse momentum squared $\langle q^2_\perp \rangle_{\rm med}$
per unit path length $\lambda$~\cite{Baier:1996sk, Salgado:2003gb,
  dEnterria:2009xfs}.
As a further simplification Q-\pythia assumes $\hat{q}$ to be time-independent which corresponds to a static uniform QGP medium.
As a consequence, the transport coefficient $\hat{q}$ becomes the only free non-perturbative
parameter within the Q-\pythia model. In what follows we will assume its value to be given by 
\begin{equation}
  \hat{q}=2\,\mathrm{GeV^2/fm}\,.
\end{equation}
More information on the way Q-\pythia evaluates the splitting kernels in Eq.~\eqref{eq:medium_splitting_func}
can be found in Refs.~\cite{Baier:1996kr, Baier:2000mf, Polosa:2006hb, Armesto:2008qh}.

Another 
MC generator to account for QGP medium effects is \jewel
(Jet Evolution With Energy Loss)~\cite{Zapp:2012ak, Zapp:2013vla, Zapp:2013zya}.
Similar to Q-\pythia, the approach of \jewel is based upon modification of the
\pythia6 parton-shower model by including $2\rightarrow2$ rescatterings between
shower partons and additional ``external'' partons representing QGP scattering
centers.
These rescatterings change the virtuality of the shower partons which afterwards
may continue to participate in the parton-shower evolution~\cite{Zapp:2013vla,
  Zapp:2012ak, Zapp:2011ek, Zapp:2008gi}.
Apart from that, \jewel also contains a Monte Carlo model for the  suppression of
gluon spectra at energy scales smaller than $w_c = \hat{q} L^2 / 2$ caused
by destructive interference in elastic scattering between emitted gluons and the QGP
medium also known as non-abelian analog of the Landau--Pomeranchuk--Migdal (LPM)
\cite{Landau:1953um, Migdal:1956tc, Gyulassy:1993hr, Zapp:2008af} effect. 
The \jewel approach to model medium effects is fully microscopic and the only
required assumption on QGP properties is the density of scattering centers which
can be evaluated using different models.
The default medium model of \jewel~\cite{Zapp:2005kt, Zapp:2012ak} is based upon
the Bjorken model of an ideal quark--gluon gas~\cite{Bjorken:1982qr} and some
Glauber calculations~\cite{Eskola:1988yh}. For definiteness, in what follows we assume
for the medium temperature a value of $\mathrm{T} = 0.55\,\text{fm}$. Furthermore,
the overlap between  colliding nuclei in the Glauber calculations is limited to the
centrality class $[0, 10]\%$.

Finally, we would like to note that the overall particle multiplicity in AA collisions
is typically about hundred times larger than in $\rm pp$ collisions.
Such drastic increase is due to the production of additional soft particles via multiple
nucleon--nucleon interactions and, presumably, the hadronization of the QGP medium.
Therefore, the presence of such additional soft background in AA collisions may require
the application of event-wide background removal techniques other than \softdrop grooming
we were using in the previous sections. This issue will be addressed below.

\subsection*{Hadron-level predictions for jet angularities in AA collisions at RHIC}

\begin{figure}
  \centering
  \includegraphics[width=0.99\linewidth]{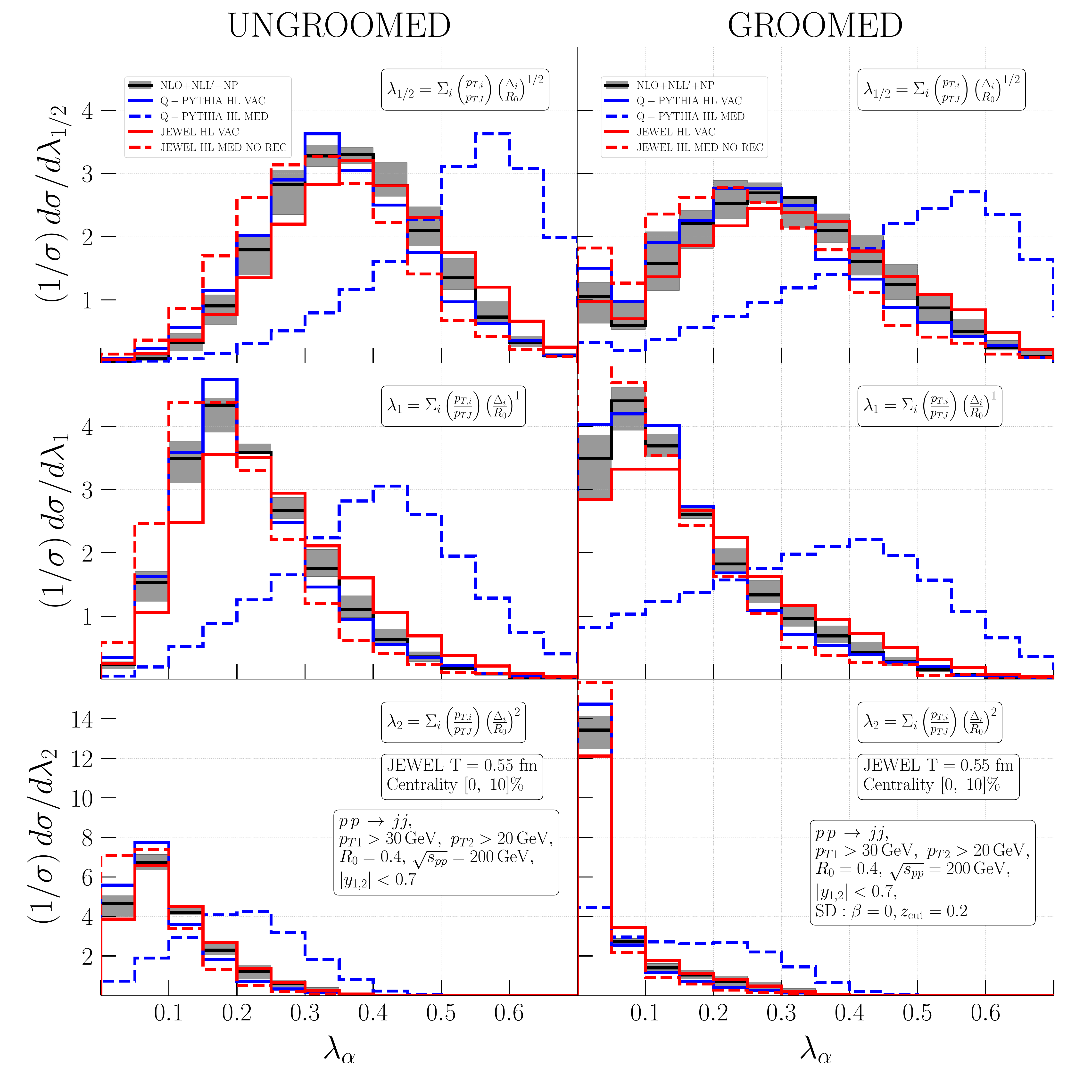}
  \centering
  \caption{Q-\pythia and \jewel predictions for jet-angularity distributions in
    $\rm pp$ (solid lines) and $\mathrm{AA}$ (dashed lines) collisions at RHIC, without physical soft-background
    subtraction. Furthermore, the hadron-level corrected resummed predictions of $\NLOpNLLp$
    accuracy are shown in black.}
  \label{fig:ang_medium_no_BGS}
\end{figure}

\begin{figure}
  \centering
  \includegraphics[width=0.99\linewidth]{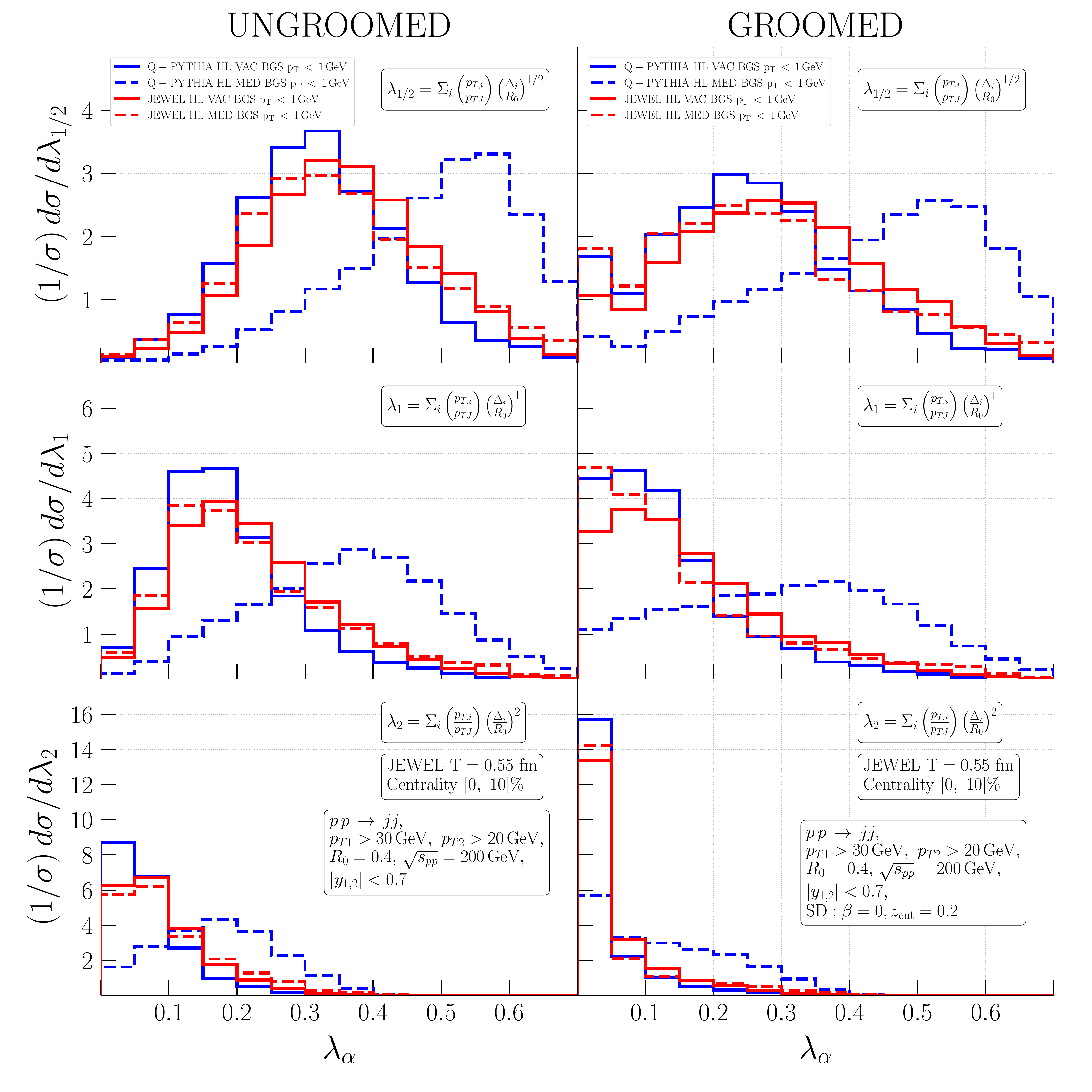}
  \centering
  \caption{Q-\pythia and \jewel predictions for jet-angularity distributions in
    $\rm pp$ (solid lines) and $\mathrm{AA}$ (dashed lines) collisions at RHIC, with soft-background subtraction,
    \emph{i.e.}\ after removal of soft particles with \mbox{$p_{T} < 1\,\text{GeV}$}
    from the final state. }
\label{fig:ang_medium_with_BGS}
\end{figure}

We present our results for the three angularities $\lambda_\alpha$ where $\alpha\in\{1/2,1,2\}$
without and including the subtraction of soft-background particles in Figs.~\ref{fig:ang_medium_no_BGS}
and \ref{fig:ang_medium_with_BGS}, respectively.
Fig.~\ref{fig:ang_medium_no_BGS} contains vacuum-level predictions from Q-\pythia (blue solid
line) and \jewel (red solid line), corresponding to the expectations for $\rm pp$ collisions.
For comparison, we also include our $\NLOpNLLp$ predictions corrected for non-perturbative
effects as presented in Section~\ref{sec:np_effects}. These results get contrasted with Q-\pythia (blue dashed line)
and \jewel (red dashed line) simulations taking into account medium effects. 
The medium Q-\pythia result does not contain particles from the bulk, soft background, whereas 
the medium \jewel result by default contains such soft particles. 
These soft background particles can easily wash out the parton-shower emission pattern. 
In Fig.~\ref{fig:ang_medium_no_BGS} we do not consider any soft-background removal measures which can be physically implemented in experiments. 
Instead, in order to evaluate the intrinsic modification of 
jets without the background contamination, we provide the medium \jewel result without the ``recoil particles" 
(red dashed line) by using an internal particle tag for medium particles in the \jewel simulation. 
In reality, such recoil particles are included in jet reconstruction and can not be physically separated from the jet. 
Therefore this ``NO REC" \jewel result should only represent an ``ideally" background subtracted result 
and is meant to show qualitative features of \jewel medium effects. 
 For the
\softdrop groomer we employed the same parameters as in Sections~\ref{sec:resum} and~\ref{sec:np_effects}.

We observe that the vacuum-level predictions obtained with \jewel and Q-\pythia wrap
the $\NLOpNLLp+\mathrm{NP}$ results, with \jewel predicting somewhat broader, Q-\pythia
somewhat narrower spectra. While the agreement among these two LO simulations is certainly not
perfect, differences are mild when considering that no efforts were made to enforce a tuned
comparison, but rather the default settings of both generators have been employed here.  
Although both codes are based upon modified versions of \pythia6, their setups for
$\rm pp$ vacuum simulations are not identical. For example, Q-\pythia employs the 
ATLAS MC09 tune~\cite{ATLAS:2010zyu}, the CTEQ 5L LO PDFs~\cite{Lai:1999wy}, and
is based on \pythia 6.4.18. \jewel, in contrast, uses the CT14 NLO PDFs~\cite{Dulat:2015mca}
and is derived from \pythia 6.4.25. While a more detailed analysis of the origin
of the deviations between the two generator predictions is interesting, this is
considered beyond the scope of the present paper. We envisage a dedicated measurement
of the considered jet angularities in $\rm pp$ collisions at sPHENIX that could then
eventually be used to align parameter choices in the simulations to obtain a satisfactory
description of the $\rm pp$-baseline results.

The AA simulation results shown in Fig.~\ref{fig:ang_medium_no_BGS} demonstrate
significant differences compared to the vacuum-level predictions just discussed.
We find that both Q-\pythia and \jewel with QGP medium effects being turned on
predict jet-angularity distributions significantly different from the ``reference''
vacuum-level results, especially for the Q-\pythia result. 
It is apparent that \jewel and Q-\pythia, while providing similar results for
the vacuum case, produce quite different medium-level distributions which significantly
deviate from each other, corresponding to their different methods to account for
medium effects in AA collisions. The Q-\pythia ansatz of modifying the QCD splitting functions
preserves parton-shower unitarity and, as a consequence, the total energy of all final-state
particles produced in a given event always equals the initial
nucleon--nucleon center-of-mass energy. Accordingly, any increase in soft-particle production is
compensated for by a decrease of the momentum of the leading partons inside a jet caused by the
energy--momentum conservation constraint. 
For \jewel, the medium effect comes from jet particles scattering with medium particles with momentum transfer. 
These medium particles can be traced as recoil particles after collisions. 
Qualitatively, in the Q-\pythia model the medium greatly increases the values of jet angularities even for \softdrop groomed jets. 
This may imply that the modified splitting functions can significantly affect hard branching kinematics. 
In contrast, the \jewel result shows that the medium decreases the values of jet angularities, 
which may be caused by loss of energy and particles inside the jet. 

There are various sophisticated background-removal techniques described in the literature on
heavy-ion physics~\cite{Dremin:2007ph, KunnawalkamElayavalli:2017hxo, Milhano:2022kzx}. 
However, in this
rather qualitative study we decided to consider just a simple subtraction of final-state particles with
$p_{T} < 1\,\text{GeV}$. 
While one might expect that \softdrop grooming helps to clean up jets from the thermal background,
this technique was specifically constructed to subtract soft wide-angle radiation with respects to the
jet axis. Soft particles included in the hard branching that satisfy the \softdrop condition therefore remain in the jet. 
In Fig.~\ref{fig:ang_medium_with_BGS} we present our corresponding Q-\pythia and \jewel default (therefore including the soft recoil particles in the simulation output) results for the
jet angularities, with soft particles removed from the final state prior to jet clustering and the
application of the event-selection criteria. Note, such procedure effectively defines new observables
which do not obey the concept of infrared and collinear (IRC) safety which forms a corner stone of the
here employed \Caesar resummation technique.
Accordingly, we here do not include our  $\NLOpNLLp+\mathrm{NP}$ predictions, given it is not completely
clear how to consistently correct them for such soft-background subtraction\footnote{We remark that a possible approach to partially account for soft-background subtraction could proceed through the extraction of corresponding transfer-matrices, where at hadron-level the particle-$p_T$ threshold criterion is invoked, see for example Ref.~\cite{Baron:2020xoi} for a related discussion of charged-track based hadronic event-shape distributions.
  However, such procedure would effectively assume that all soft particles with momentum below the
  particle-$p_T$ threshold are described by non-perturbative QCD and hence can be considered separately
  from the resummed calculations. An assumption that remains to be verified.}. 

\begin{figure}
  \centering
  \includegraphics[width=0.99\linewidth]{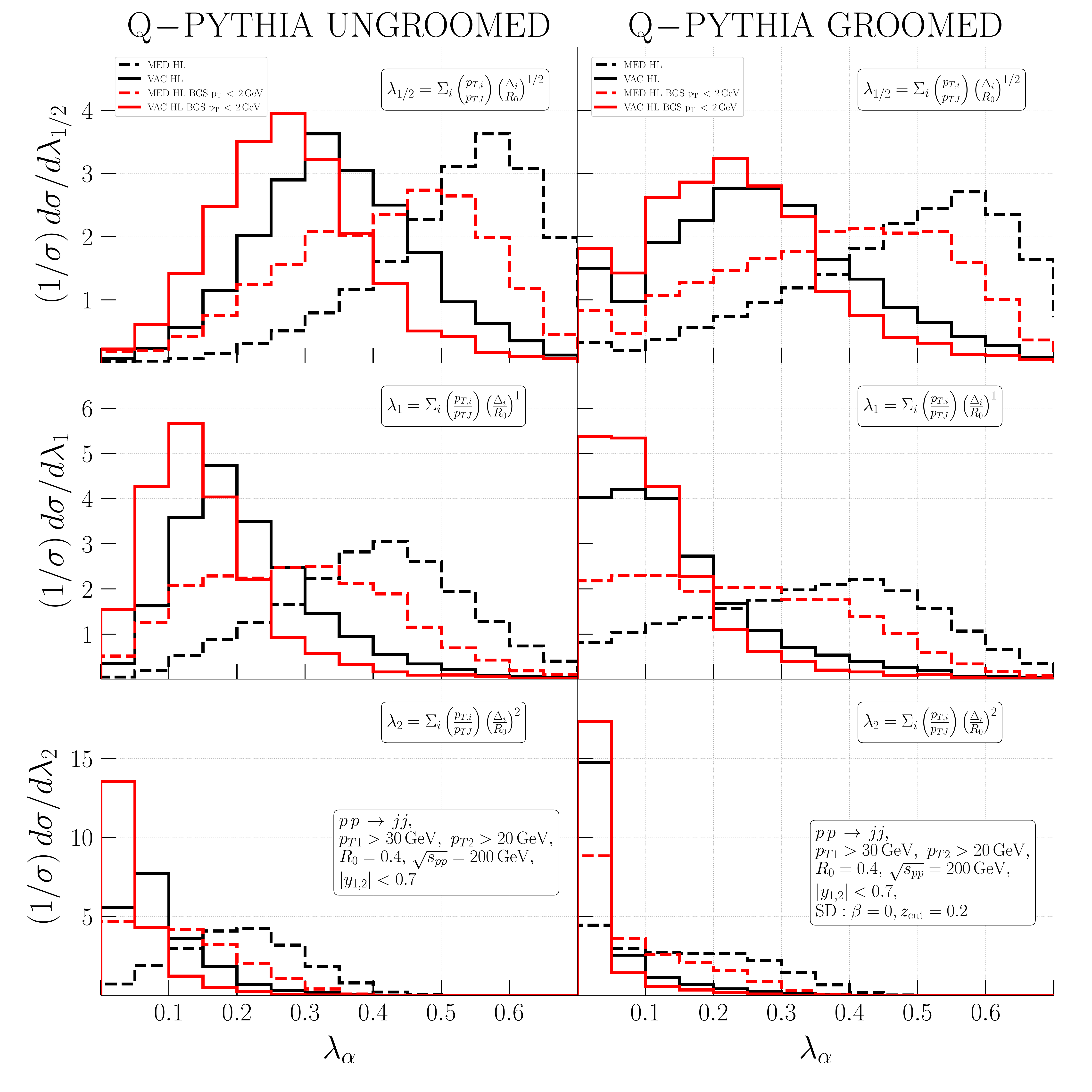}
  \centering
  \caption{Q-\pythia predictions for jet-angularity distributions in
    $\rm pp$ (solid lines) and $\mathrm{AA}$ (dashed lines) collisions at RHIC, with (red lines) and without (black lines) soft-background subtraction,
    \emph{i.e.}\ after removal of soft particles with \mbox{$p_{T} < 2\,\text{GeV}$}
    from the final state.}
\label{fig:ang_qpythia_pT_var}
\end{figure}

By comparing the results shown in Figs.~\ref{fig:ang_medium_no_BGS} and
\ref{fig:ang_medium_with_BGS} we observe that the removal of soft particles with
$p_{T} < 1$ GeV mildly affects the Q-\pythia results, both for the vacuum simulations
as well as for the AA runs. However, it effectively removes soft recoil particles in the \jewel medium-level simulations,
producing a distribution (dashed red line in Fig.~\ref{fig:ang_medium_with_BGS}) which is close to the medium-level distribution without recoil particles (dashed red line in Fig.~\ref{fig:ang_medium_no_BGS}) and also the vacuum-level distribution (solid red line in Fig.~\ref{fig:ang_medium_with_BGS}).
The stability of the Q-\pythia results with respect to the removal of soft
particles below 1 GeV can be explained by the fact that the Q-\pythia model is based on the
modification of the QCD emission pattern according to Eq.~\eqref{eq:medium_splitting_func}, thereby,
essentially, imposing a redistribution of energy between soft and hard particles produced inside
a given jet due to the interaction with the external QGP medium. 
Accordingly, soft final-state particles produced in AA simulations with Q-\pythia cannot directly
be interpreted as a thermal QCD background from overlaid nucleon--nucleon interactions that produce uniformly distributed soft particles with very low energies. 
In Fig.~\ref{fig:ang_qpythia_pT_var} we further show the Q-\pythia results with a more stringent cut to remove soft particles below 2 GeV, and 
the medium modification pattern remains significant. 

Finally, we would like to note that the \jewel AA simulations after removal of soft particles
with $p_{T} < 1$ GeV, which may include both jet and medium particles, differ marginally with the corresponding $\rm pp$ results.
The differences between the vacuum-level Q-\pythia and \jewel results is sometimes even larger
than the difference between vacuum-level and medium-level distributions from \jewel.
That is, the uncertainty of vacuum-level Monte Carlo predictions is larger than the size of the medium effect \jewel predicts. 
We therefore conclude that a detailed comparison of vacuum-level simulations against
experimental data collected at RHIC in $\rm pp$ collisions will be of utmost importance
to validate and challenge the available theoretical predictions based on the intriguing
combination of perturbative and non-perturbative QCD phenomena. Besides Monte Carlo
simulation tools this also includes analytic predictions based on all-order resummation,
corrected for hadronization and underlying event effects. Such detailed comparison will
allow us to establish a baseline for the understanding of jet substructure in $\rm pp$ collisions
at RHIC which is necessary for getting control over jet-substructure studies in AA scattering
events.

\section{Conclusions}
\label{sec:conclusions}
Jet-substructure measurements provide unique insights into the dynamics of
QCD and constitute a superb testbed for theoretical approaches. In this study
we considered the well-known class of jet-angularity observables for the case
of $\rm pp$ and $\mathrm{AA}$ collisions at RHIC. Thereby the $\rm pp$ setup serves as a rather
well understood and well controlled reference towards studying the impact of
medium-effects as anticipated from the creation of a QGP in heavy-ion collisions.

To this end, we derived $\NLOpNLLp$ accurate predictions for three distinct angularities
of QCD jets, with and without \softdrop grooming, produced in $\rm pp$ collisions corresponding to the setup of the sPHENIX
experiment currently operating at RHIC. To account for non-perturbative effects from underlying
event and hadronization, we correct our predictions with the help of transfer matrices which
have been shown to provide a better description of collider data than a simple multiplicative
reweighting approach~\cite{Reichelt:2021svh}.
Our detailed study of non-perturbative contributions demonstrated a dominant r\^{o}le of
hadronization corrections and a suppression of contributions from multi-parton interactions
compared to an earlier LHC study~\cite{Reichelt:2021svh}.
Accordingly, we claim that dedicated jet-substructure studies in $\rm pp$ collisions at RHIC can, 
to a large extend, help to disentangle and constrain hadronization effects from the underlying event
which is clearly not possible at the LHC due to the much higher beam energy.  

The $\NLOpNLLp$ resummed predictions corrected for non-perturbative results were compared against
\sherpa simulations of MC@NLO accuracy and indeed very good agreement between the two results,
taking into account theoretical uncertainties, was found. We envisage that a future measurement
of sPHENIX and the detailed comparison against our and other theoretical predictions will help
to further probe and constrain MC models for non-perturbative effects in $\rm pp$ collisions at
energies much lower than at LHC.

While we consider the $\NLOpNLLp+\mathrm{NP}$ and MC@NLO \sherpa predictions a main deliverable
of this paper, we also provided a qualitative estimate for the impact of QGP medium-effects on
jet-substructure observables. To this end, we simulated AA collisions with two 
MC
generators, namely Q-\pythia and \jewel. 
We found that the jet-angularity predictions from Q-\pythia and \jewel show a significantly
different behavior. Whereas the default \jewel setting simulates high-multiplicity soft-particle production originating
from the external QGP media, Q-\pythia modifies the parton-shower splitting kernels according to
the BDMPS-Z formalism, resulting in an increased emission of soft particles from the hard-process
final state. Accordingly, their multiplicity is limited by energy--momentum conservation,
\emph{i.e.}\ the constraint that the energies of all final-state particles have to sum up to the
nucleon--nucleon center-of-mass energy. This, as was also argued in the original Q-\pythia
publication~\cite{Armesto:2009fj}, can be seen as a limitation of the current model since in AA collisions
multiple soft particles can furthermore be produced due to the thermalization of the QGP and additional
nucleon--nucleon interactions, such that the total energy of all final-state particles can easily
exceed $\sqrt{s_{\text{NN}}}$. 
Q-\pythia predicts a significant enhancement of jet angularity values. On the other hand, by comparing the \jewel jet angularity predictions for $\rm pp$ and AA (without recoil particles) collisions, a significant decrease of jet angularity values was observed in \jewel. 
After soft-background subtraction, the jet angularity predictions for AA collisions simulated with \jewel
showed very moderate medium modifications compared to the vacuum-level results, whereas the Q-\pythia AA
results still exhibit striking differences.
Given that jet-substructure modifications observed in previous experiments do not tend to support such strong
effects compared to $\rm pp$ results~\cite{Connors:2017ptx}, we would assume that after detailed comparison against
upcoming new results from sPHENIX the Q-\pythia model may need to be revised.
We also argue that the observed marginal difference between \jewel $\rm pp$ and $\mathrm{AA}$ results after background removal, 
as compared to the difference between \jewel and Q-\pythia $\rm pp$ predictions, 
highlights the importance of precision understanding of the $\rm pp$ results in order to meaningfully extract medium modifications. 
Whereas in this paper we used a simple removal of final-state particles with $p_{T} < 1$ GeV, a more sophisticated
background-subtraction technique such as described in Refs.~\cite{Dremin:2007ph, KunnawalkamElayavalli:2017hxo, Milhano:2022kzx} should be used
when comparing Monte Carlo predictions against data. Each of these techniques, in turn, leads to somewhat differently
defined jet-substructure observables and hence requires a dedicated investigation and analytic understanding which we leave to future studies.

\section*{Acknowledgments}
SS acknowledges support from BMBF (05H21MGCAB) and from Deutsche Forschungsgemeinschaft
(DFG, German Research Foundation) — project number 456104544. 
The work of DR was supported by the STFC IPPP grant (ST/T001011/1).
The work of OF and YTC is supported in part by the US Department of Energy (DOE) Contract No.~DE-AC05-06OR23177, under which Jefferson Science Associates, LLC operates Jefferson Lab, and by the Department of Energy Early Career Award grant DE-SC0023304.
We are grateful to Liliana Apolin\'{a}rio  and N\'{e}stor Armesto for providing the version of the Q-Pythia code specifically tuned for the RHIC setup.
We also would like to thank Roli Esha and Megan Connors for useful and fruitful discussions on the potential of the sPHENIX experiment for jet angularity measurements. 
Most of the simulation is conducted with computing facilities of the Galileo cluster at the Department of Physics and Astronomy of Georgia State University. 

\appendix
\renewcommand{\thesection}{\Alph{section}} %
\section{Connection between transfer matrices and shape functions}\label{app:shapefuncs}
The non-perturbative power corrections of general infrared and collinear (IRC) safe jet-substructure observables may be included by convolving the resummed, perturbative calculations with the corresponding shape functions originally introduced by Korchemsky and Sterman \cite{Korchemsky:1999kt}. For jet angularities, the convolution may take the following form \cite{Kang:2018vgn, Yan:2023xsd} if the non-perturbative correction is naively modeled by a single-scale contribution\footnote{The non-perturbative corrections to \softdrop jet mass was studied in details in \cite{Hoang:2019ceu, Ferdinand:2023vaf}. There the shape function is involved in a non-trivial convolution caused by the \softdrop procedure.},
\begin{equation}
	\frac{d\sigma}{d\lambda_{\alpha}^{\text{HL}}}
	=\int d\epsilon~f(\epsilon) \int d\lambda_\alpha^{\text{PL}}\frac{d\sigma}{d\lambda_\alpha^{\text{PL}}}\delta\Big(\lambda_\alpha^{\text{HL}}-\lambda_\alpha^{\text{PL}}-C_{\alpha}^{\beta,\zc}\epsilon^{\gamma_\alpha^\beta}\Big)\;,
\end{equation}
where the function $f(\epsilon)$ is the shape function of the jet angularity $\lambda_\alpha$. Its functional argument $\epsilon$ is the low energy contribution to $\lambda_\alpha$ from the non-perturbative soft momenta. Here we suppress the dependence of the observable on the soft-drop parameters $\beta$ and $\zc$ for simplicity. Because of the additivity of jet angularities from individual jet-constituent momenta, the value of the observable at hadron level ($\lambda_\alpha^{\text{HL}}$) is offset from the value at parton level ($\lambda_\alpha^{\text{PL}}$) by an amount $C_{\alpha}^{\beta,\zc}\epsilon^{\gamma_\alpha^\beta}$. The coefficient $C_{\alpha}^{\beta,\zc}$ and the power $\gamma_\alpha^\beta$ in general depend on the angularity and soft-drop parameters. Here we can use a power-counting argument to derive the dependence on these parameters. The measurement of $\lambda_\alpha$ constrains the values of $z$ and $\Delta$ of a particle emission,
\begin{equation}
	\lambda_\alpha\sim z\Big(\frac{\Delta}{R_0}\Big)^\alpha\;.
\end{equation}
On the other hand, the \softdrop procedure induces another constraint through the \softdrop condition,
\begin{equation}
	z\sim \zc \Big(\frac{\Delta}{R_0}\Big)^\beta\;.
\end{equation}
These two constrains then determine the values of $z$ and $\Delta$,
\begin{equation}
	z\sim \zc \Big(\frac{\lambda_\alpha}{\zc}\Big)^{\frac{\beta}{\alpha+\beta}},~~~~~~~~\Delta\sim R_0\Big(\frac{\lambda_\alpha}{\zc}\Big)^{\frac{1}{\alpha+\beta}}\;,
\end{equation}
which induces the following characteristic energy scale $\epsilon$,
\begin{equation}
	\epsilon\sim p_Tz\Delta\sim p_T \zc R_0\Big(\frac{\lambda_\alpha}{\zc}\Big)^{\frac{1+\beta}{\alpha+\beta}}\;.
\end{equation}
Non-perturbative soft physics at this scale may contribute to the angularity value by an amount $\delta \lambda_\alpha$, where
\begin{equation}
	\delta \lambda_\alpha\sim \zc \Big(\frac{\epsilon}{p_T \zc R_0}\Big)^{\frac{\alpha+\beta}{1+\beta}}\;.
\end{equation}
In this case $C_{\alpha}^{\beta,\zc} \sim \zc \Big(1/p_T \zc R_0\Big)^{\frac{\alpha+\beta}{1+\beta}}$ and $\gamma_\alpha^\beta \sim \frac{\alpha+\beta}{1+\beta}$.

\begin{figure}
  \centering
  \includegraphics[width=0.47\linewidth]{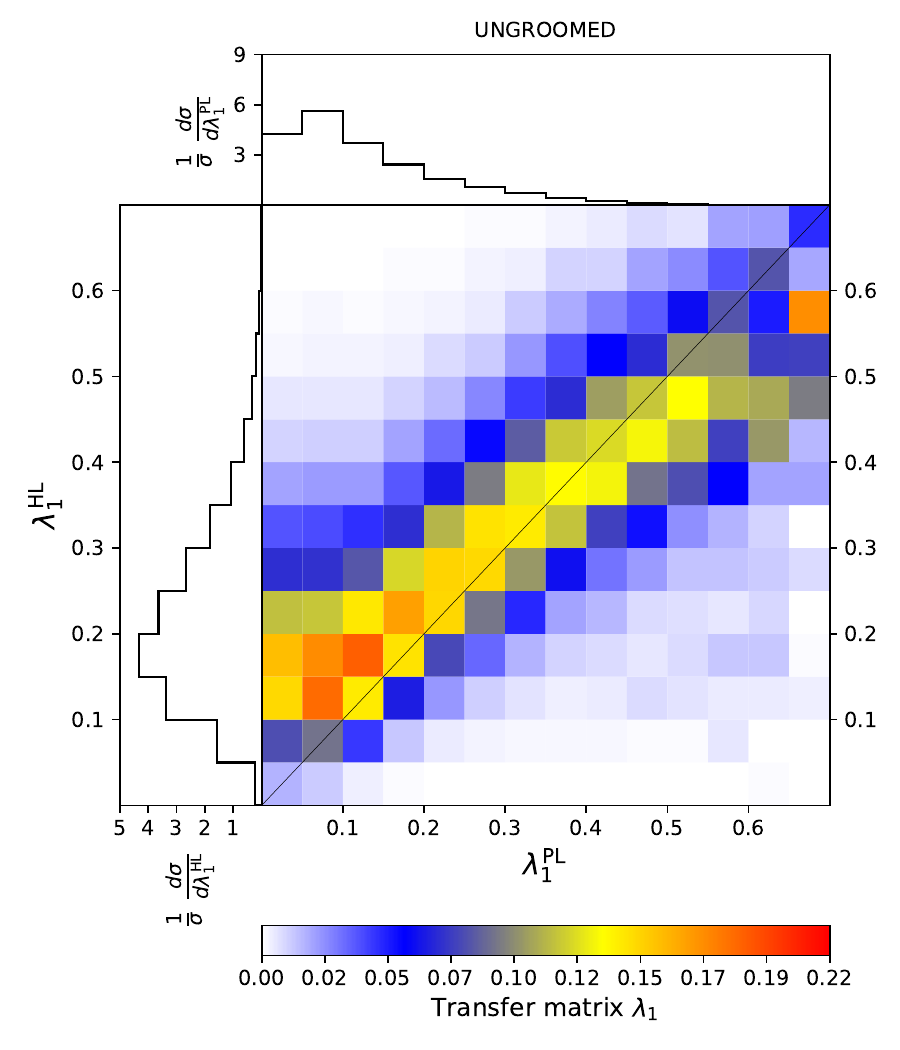} 
  \hspace{1em}
  \includegraphics[width=0.47\linewidth]{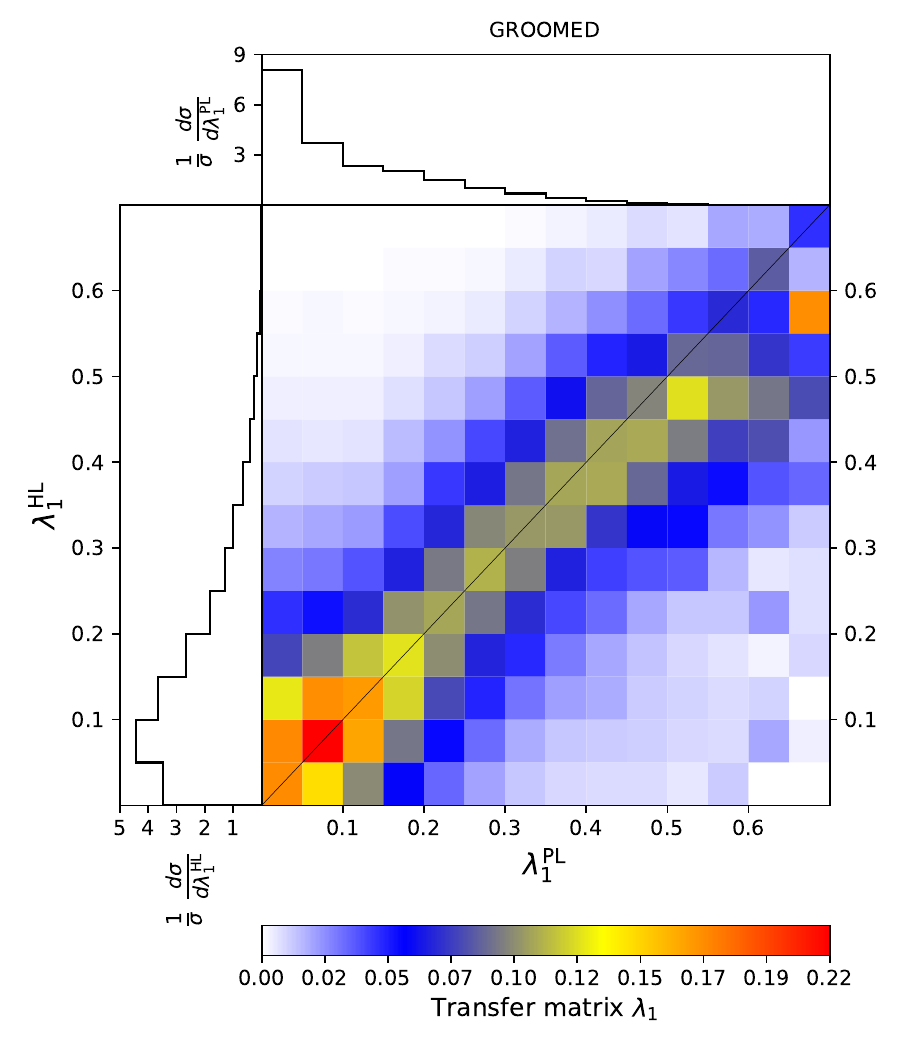}
  \centering
  \caption{Transfer matrices defined according to Eq.~\eqref{eq:Tfinal} for
    the ungroomed (left) and groomed (right) angularity $\lambda_{1}$, extracted from
    \sherpa MC@NLO hadron-level simulations for the default hadronisation tune. }
\label{fig:TM_Sherpa_width}
\end{figure}

\begin{figure}
  \centering
  \includegraphics[width=0.47\linewidth]{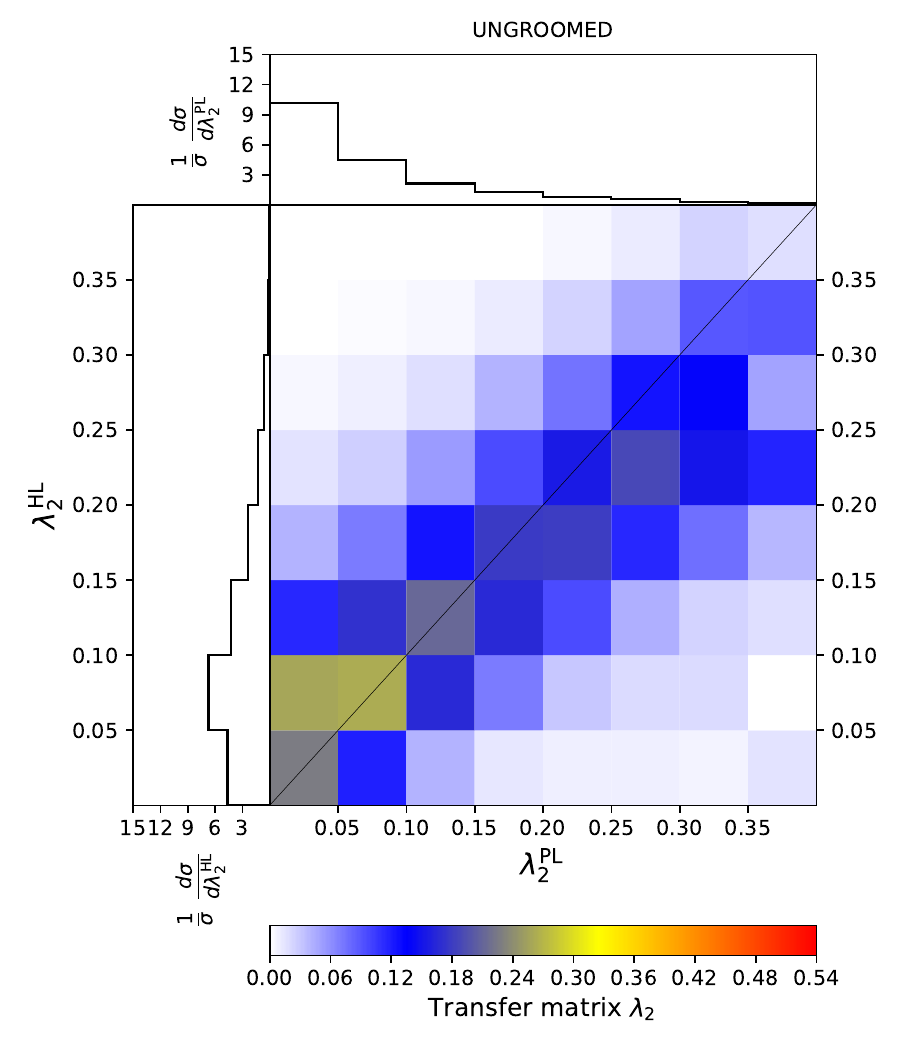} 
  \hspace{1em}
  \includegraphics[width=0.47\linewidth]{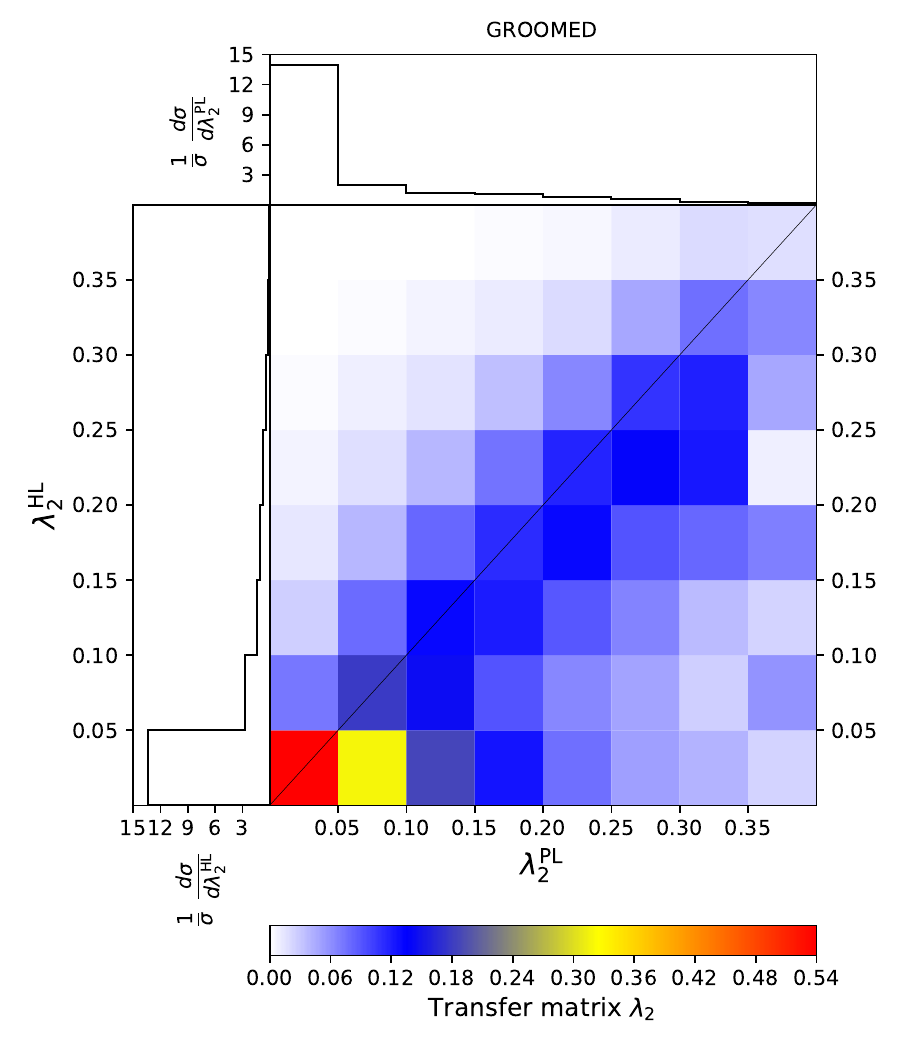}
  \centering
  \caption{Transfer matrices defined according to Eq.~\eqref{eq:Tfinal} for
    the ungroomed (left) and groomed (right) angularity $\lambda_{2}$, extracted from
    \sherpa MC@NLO hadron-level simulations for the default hadronisation tune. }
\label{fig:TM_Sherpa_thrust}
\end{figure}

After integrating out the soft momentum variable $\epsilon$, the hadronic cross section can be expressed in the following integral form,
\begin{align}
	\frac{d\sigma}{d\lambda_{\alpha}^{\text{HL}}}
	=&\int d\lambda_\alpha^{\text{PL}}\frac{d\sigma}{d\lambda_\alpha^{\text{PL}}}\left[\frac{1}{C_{\alpha}^{\beta,\zc}\gamma_\alpha^\beta}\left(\frac{\lambda_\alpha^{\text{HL}}-\lambda_\alpha^{\text{PL}}}{C_{\alpha}^{\beta,\zc}}\right)^{\frac{1}{\gamma_\alpha^\beta}-1} f\left(\left(\frac{\lambda_\alpha^{\text{HL}}-\lambda_\alpha^{\text{PL}}}{C_{\alpha}^{\beta,\zc}}\right)^{\frac{1}{\gamma_\alpha^\beta}}\right) \right]\\
        =&\int d\lambda_\alpha^{\text{PL}}\frac{d\sigma}{d\lambda_\alpha^{\text{PL}}}\mathcal{T}\left(\lambda_\alpha^{\text{HL}}|\lambda_\alpha^{\text{PL}}\right)\;,
\end{align}
where the term in the square brackets can be identified as the transfer matrix $\mathcal{T}(\lambda_\alpha^{\text{HL}}|\lambda_\alpha^{\text{PL}})$ interpreted as the conditional probability of $\lambda_\alpha^{\text{HL}}$, given the value of $\lambda_\alpha^{\text{PL}}$ before the non-perturbative process. The naive convolution of the shape function implies that the transfer matrix depends only on $\lambda_\alpha^{\text{HL}}-\lambda_\alpha^{\text{PL}}$. The extracted transfer matrices in our work roughly show a consistent trend with equal probability along the diagonal lines of constant $\lambda_\alpha^{\text{HL}}-\lambda_\alpha^{\text{PL}}$ in a restricted region of angularity, see Fig.~\ref{fig:TM_Sherpa} in Section~\ref{sec:np_effects} and Figs.~\ref{fig:TM_Sherpa_width} and \ref{fig:TM_Sherpa_thrust} provided in this appendix. By projecting the transfer matrix onto a constant $\lambda_\alpha^{\text{HL}}+\lambda_\alpha^{\text{PL}}$ line we may extract the shape function consistent with the Monte Carlo non-perturbative correction. 
In fact, the extracted transfer matrix does not follow the scaling of $\lambda_\alpha^{\text{HL}}-\lambda_\alpha^{\text{PL}}$ especially in the large or small angularity regions. Therefore the transfer matrix can be a general approach to quantify non-perturbative corrections.

\clearpage

\phantomsection
\addcontentsline{toc}{section}{References}
\bibliographystyle{jhep}
\bibliography{references}

\end{document}